\documentclass[useAMS,usenatbib]{mnras}
\usepackage{rotating}
\usepackage{lscape}
\usepackage{longtable}
\usepackage{lastpage}
\usepackage{graphicx}
\usepackage{xcolor}
\usepackage{amssymb}

\def\ms{ms$^{-1}$}

\def\me{M$_{\rm{\oplus}}$}

\def\gccc{gcm$^{-3}$}
\def\re{R$_{\rm{\oplus}}$}

\title{GJ~357: A low-mass planetary system uncovered by precision radial-velocities and dynamical simulations}
\author[J.~S. Jenkins et al.]{J.~S. Jenkins$^{1,2}$\thanks{E-mail: jjenkins@das.uchile.cl}, F.~J. Pozuelos$^{3,4,5}$, M. Tuomi$^{1,6}$ Z.~M. Berdi\~nas$^1$, M.~R. D\'iaz$^1$,  
\newauthor 
J.~I. Vines$^1$, Juan C. Su\'arez$^{5,7}$, P.~A. Pe\~na Rojas$^1$\\
$^1$Departamento de Astronom\'ia, Universidad de Chile, Camino El Observatorio 1515, Las Condes, Santiago, Chile, Casilla 36-D\\
$^2$Centro de Astrof\'isica y Tecnolog\'ias Afines (CATA), Casilla 36-D, Santiago, Chile\\
$^3$Space sciences, Technologies and Astrophysics Research (STAR) Institute, Universite de Liege, 19C Allee du 6 Aout, B-4000 Liege, Belgium\\
$^4$EXOTIC Lab, UR Astrobiology, AGO Department, University of Li\`ege, 4000 Li\`ege, Belgium\\ 
$^5$Dpt. F\'isica Te\'orica y del Cosmos, Universidad de Granada, Campus de Fuentenueva s/n, 18071, Granada, Spain\\
$^6$Center for Astrophysics, University of Hertfordshire, College Lane Campus, Hatfield, Hertfordshire, UK, AL10 9AB\\
$^7$Instituto de Astrof\'isica de Andaluc\'ia (CSIC), Glorieta de la Astronom\'ia s/n, 18008, Granada, Spain
}
\begin{document}

\date{Accepted 15th of October, 2019}

\pagerange{\pageref{firstpage}--\pageref{lastpage}} \pubyear{2019}

\maketitle

\label{firstpage}

\begin{abstract}

We report the detection of a new planetary system orbiting the nearby M2.5V star GJ~357, using precision radial-velocities from three separate echelle spectrographs, HARPS, HiRES, and UVES.  Three small planets have been confirmed in the system, with periods of 9.125$\pm$0.001, 3.9306$\pm$0.0003, and 55.70$\pm$0.05~days, and minimum masses of 3.33$\pm$0.48, 2.09$\pm$0.32, and 6.72$\pm$0.94~M$_{\oplus}$, respectively.  The second planet in our system, GJ~357c, was recently shown to transit by the Transiting Exoplanet Survey Satellite \citep[TESS;][]{luque19}, but we could find no transit signatures for the other two planets.  Dynamical analysis reveals the system is likely to be close to coplanar, is stable on Myrs timescales, and places strong upper limits on the masses of the two non-transiting planets b and d of 4.25 and 11.20~M$_{\oplus}$, respectively.  Therefore, we confirm the system contains at least two super-Earths, and either a third super-Earth or mini-Neptune planet.  GJ~357b\, \&\, c are found to be close to a 7:3 mean motion resonance, however no libration of the orbital parameters was found in our simulations.  Analysis of the photometric lightcurve of the star from the TESS, when combined with our radial-velocities, reveal GJ~357c has an absolute mass, radius, and density of $2.248^{+0.117}_{-0.120}$~\me, $1.167^{+0.037}_{-0.036}$~\re, and $7.757^{+0.889}_{-0.789}$~\gccc, respectively.  Comparison to super-Earth structure models reveals the planet is likely an iron dominated world.  The GJ~357 system adds to the small sample of low-mass planetary systems with well constrained masses, and further observational and dynamical follow-up is warranted to better understand the overall population of small multi-planet systems in the solar neighbourhood.

\end{abstract}


\section{Introduction}

In recent years, M dwarf planetary systems have shown themselves to exhibit a wide range in diversity, whilst also providing a rich hunting ground for instruments and teams with the capability to scrutinise them. The high impact discoveries of Proxima Centauri~b \citep{anglada-escude16} and Barnard's Star~b \citep{ribas18} highlight this. The first planets found orbiting these stars through radial-velocity (RV) measurements, started to hint at significant differences when compared to the populations of exoplanets that were emerging from the studies of Sun-like stars. For example, it was quickly observed that there appeared a relative lack of short-period gas giant planets \citep{bonfils05a,johnson10}. This finding was explained as evidence for the core accretion model of planet formation, whereby large planets are more difficult to form orbiting small stars as they were likely orbited initially by much smaller proto-planetary disks \citep{laughlin04a}. Yet, some features were also in agreement with studies of planets orbiting more massive stars. Gas giant planets have been found in higher abundance orbiting Sun-like stars \citep{fischer05,adibekyan12,jenkins17}, and this bias appears to hold down into the M dwarf regime also \citep{schlaufman10,rojas-ayala10,montes18}.

One key feature that spurred the early efforts to search for planets orbiting M stars, was the fact that the biases work in favour of smaller planet detections. These stars could represent a pathway to studying the population of planets with masses below that of Neptune. Indeed, small planets began to appear in abundance orbiting M dwarfs \cite[e.g.][]{bonfils05a,forveille09}, and furthermore, it was witnessed that multi-planet systems were the norm, with a range of dynamically packed configurations beginning to appear \citep[e.g. GJ~581, GJ~667C:][]{mayor09b,anglada-escude13}. Large RV surveys of M stars have revealed that the planet fraction is around 100\% \citep{bonfils2013,tuomi14b}. Furthermore, the occurrence rate of planets in these systems is also high, particularly in the stellar habitable zone, with a rate of 0.21$^{+0.03}_{-0.05}$ planets per star with masses between 3$\--$10~\me \citep{tuomi14b}.

More recently, transit surveys have also turned towards M stars as prime targets in their quest for ever smaller exoplanets. The transit bias works in a similar manner to the RV bias, such that small stars allow the detection of smaller planets as the transit dip has a high signal-to-noise ratio. TRAPPIST-1 \citep{gillon17} was a prime example of the type of planet systems that small M stars can host. Transits from seven small rocky planets on short period orbits were found, allowing detailed studies to be conducted on the possible atmospheres and their constituents of small planets orbiting a single star, and hence with similar initial conditions \citep[e.g.][]{alberti17,ducrot18,moran18,miles-paez19,burdanov19}.  

Recent, instrumentation in space has allowed an even deeper study of the population of planets orbiting small stars.  The Kepler Space Telescope \citep{borucki10} observed thousands of M stars in its viewing zone, and revealed similar statistical constraints to that of the RV studies.  \citet{kopparapu13} used their own updated prescriptions for the extent of the conservative and optimistic habitable zones, to reveal occurrence rates of 0.48$^{+0.12}_{-0.24}$ and 0.53$^{+0.08}_{-0.17}$ planets per star, respectively. \citet{dressing15} found 159 planet candidates, revealing an occurrence rate of 2.5$\pm$0.2 planets per star with radii 1$\--$4~$R_{\oplus}$ and orbital periods less than 200 days. They also found a higher rate for the smallest planets, but in radius not mass. Within orbital periods of 50 days, planets with radii $1\--1.5~R_{\oplus}$ have an occurrence rate of 0.56$^{+0.06}_{-0.05}$ planets per star, whereas those with radii $1.5\--2.0~R_{\oplus}$ have a rate of 0.46$^{+0.07}_{-0.05}$ planets per M star.

The latest space mission that will study a wider sample of M stars is the Transiting Exoplanet Survey Satellite (TESS) project \citep{TESS}. According to simulations, TESS will find $990\pm350$ planets among $715\pm255$ early-to-mid M dwarf host stars \citep{ballard19}. One key goal of TESS, a Level 1 science goal, is to provide constrained radii and masses for 50 planets below 4$R_{\oplus}$.  In this work, we discuss a rich planetary system orbiting the M star GJ~357, with one such candidate found to transit it's star \citep[see][]{luque19}. The mass and radius of this transit candidate makes the system a TESS Level 1 science target. GJ~357 is located within 20~pc of the Sun, and is an early-M star that is on the main sequence. The star has a mass and radius of $0.378 \pm 0.03~M_{\odot}$ and $0.359 \pm 0.011~R_{\odot}$, respectively, and is found to be inactive and metal-poor ([Fe/H] = -0.30 $\pm$ 0.09~dex). The main stellar parameters that we are interested in for this work are listed in Table~\ref{tab:star}.

\begin{table}
    \centering
    \caption{Some key stellar parameters for GJ~357.}\label{tab:star} 
    \begin{tabular}{ccc}
    \hline
    Parameter     & Value & Source \\
     \hline    
    TESS Names & TIC413248763 (TOI-562) &  \\
    RA [hr:min:secs] & 09:36:01.63725 & Gaia \\
    Dec [deg:min:sec] & -21:39:38.87828 & Gaia \\
    Parallax [mas] & 105.8830 $\pm$ 0.0569 & Gaia \\
    Distance [pc] & 19.5485 $\pm$ 0.0106 & Gaia \\
    $V$ [mag] & 10.91 $\pm$ 0.02 & Hipparcos \\
    SpT & M2.5V & N14 \\
    T$_{\rm eff}$ [K]    & $3344 \pm 110$     & N14\\
    log$g$ [cms$^{-2}$]    & $4.906 \pm 0.062$  & St18\\
    $\left[\rm{Fe/H}\right]$	 [dex]    & $-0.30 \pm 0.09$     & N14\\
    $L/L_{\odot}$ & 0.0175$\pm$0.0011 & St18 \\
    $R~[R_{\odot}$] & 0.359$\pm$0.011 & Mu18\\
    $M~[M_{\odot}$] & 0.378$\pm$0.03 & Mu18\\
    $v~sin(i)$ [kms$^{-1}$] & 1.21 & H17\\
    $P/sin(i)$ [days] & 18.7 $\pm$ 4.3 & H17\\
    $logR'_{HK}$ [dex] & -5.652 & H17\\ \hline
    \end{tabular}
    % \hline
    Gaia: \citet{gaia16,gaia18}; Hipparcos: \citet{perryman97}; H17: \citet{houdebine17}; Mu18: \citet{muirhead18}; N14: \citet{neves14}; St18: \citet{stassun18} .
\end{table}

The paper is formatted as follows: in $\S$~\ref{rv_cands} we discuss the RV detection of our planetary system, along with the properties of the host star. In $\S$~\ref{act} we discuss the stellar activity analysis we performed to rule out activity as the source of the signals, whilst in $\S$~\ref{tess} we show the TESS light curve and discuss the joint model that provides our overall constraints on the planetary system. Finally, in $\S$~\ref{sum} we discuss the impact of this result and summarise our findings.

\section{RV candidates}\label{rv_cands}

GJ~357 was a target star in the \citet{zechmeister2009}, \citet{bonfils2013}, \citet{tuomi14b} and \citet{tuomi2019} samples studying the RV variability of nearby M dwarfs. \citet{tuomi2019} reported three candidate planets orbiting the star based on an analysis of combined HARPS, HiRES, and UVES radial velocities. We revisit the detections of the corresponding signals in this section and the data analysed in \citeauthor{tuomi2019} is referred to as the ``old data".

As discussed by \citet{feng2017}, RVs calculated for independent spectrograph orders can be used as independent data sets. We have used the HARPS RVs for each of the independent 72 orders to calculate the radial velocity data sets as in \citet{AngladaButler2012}, but we have also obtained independent velocity sets by dividing the 72 HARPS orders into three subsets of 24 orders, and by neglecting the first such subset because it corresponds to the noisiest velocities calculated for the bluest wavelengths \citep{AngladaButler2012}. Consequently, we have two data sets obtained for HARPS orders 25-48 and 49-72 that can be used as independent RV data sets in the analyses. Moreover, we have analysed these sets in combination with data from other instruments, including a new reduction of UVES velocities, and accounted for the wavelength-dependent variability in these sets by using the so-called differential velocities as noise proxies \citep{feng2017}. We refer to this data as the ``new data".

The combined RV data of GJ~357 (old and new) was analysed by applying the delayed-rejection adaptive Metropolis (DRAM) Markov chain Monte Carlo (MCMC) algorithm \citep{haario01,haario06} that is a generalised version of the Metropolis-Hastings posterior sampling algorithm \citep{metropolis1953,hastings1970} and has been applied in e.g. \citet{butler2017}, \citet{diaz2018}, and \citet{tuomi2018}. This sampling technique was applied in two steps \citep[see e.g.][]{tuomi2019}. First, for a model with $k \geq 1$ Keplerian signals, we performed searches for periodic signals by obtaining a sample from the posterior probability density in order to identify what signal periods correspond to the highest probability maxima in this density. In the second step, we obtained statistically representative samples from the joint posterior density of model parameters by starting the corresponding samplings at or near the highest probability maximum in the period space. The obtained results were then subjected to statistical tests in order to determine whether the identified probability maxima corresponded to statistically significant signals \citep[see e.g.][]{tuomi2019}. We calculated the Bayesian information criterion for each model with different numbers of Keplerian signals to estimate the significances of signals because of its simplicity \citep{liddle2007} and robustness in practice \citep{feng2016}.

\begin{figure*}
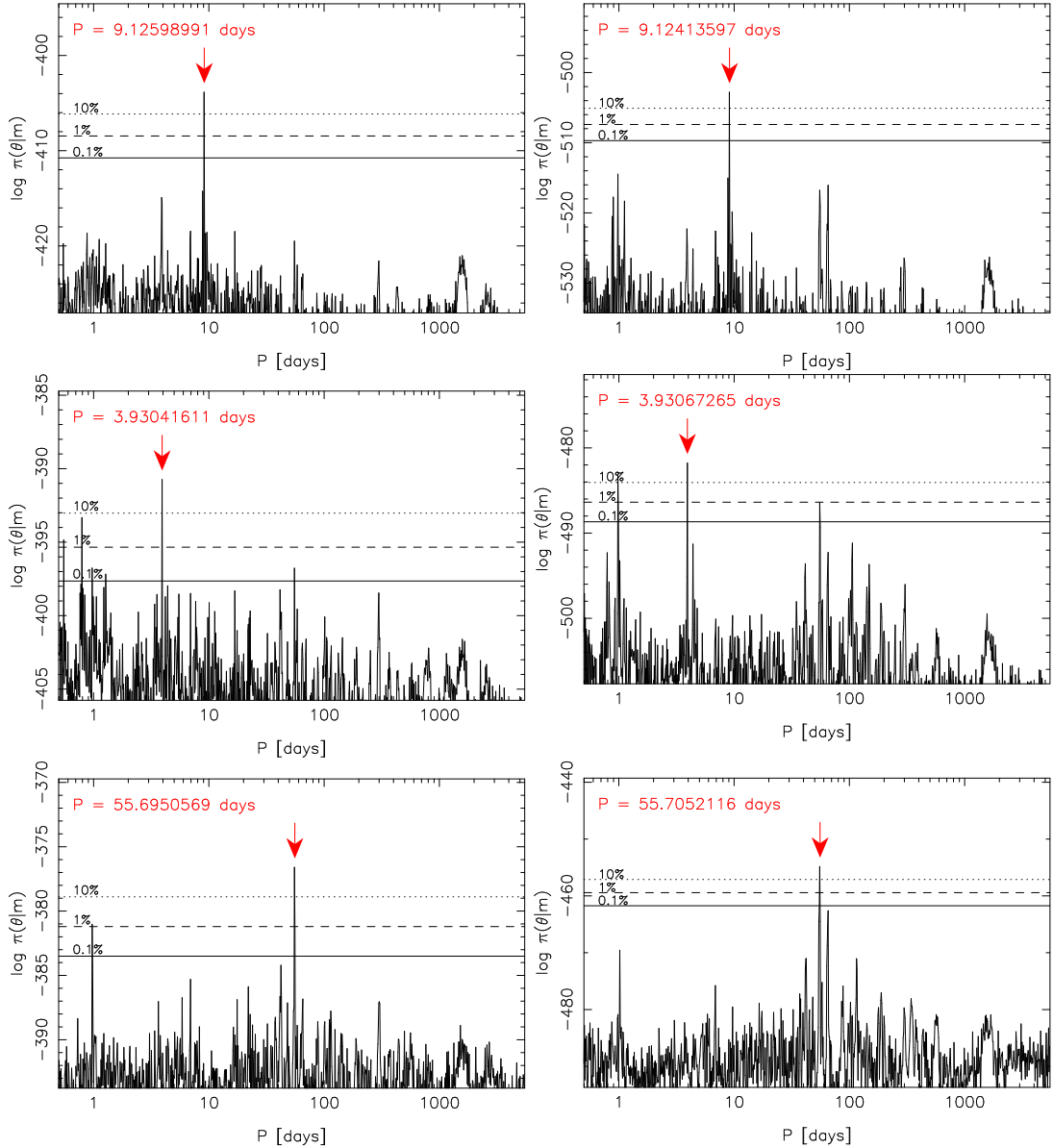

\center
\includegraphics[angle=270, width=0.40\textwidth,clip]{rv_GJ357_OLD_01_pcurve_b.ps}
\includegraphics[angle=270, width=0.40\textwidth,clip]{rv_GJ357_3AP_ALL_01_pcurve_b.ps}

\includegraphics[angle=270, width=0.40\textwidth,clip]{rv_GJ357_OLD_02_pcurve_c.ps}
\includegraphics[angle=270, width=0.40\textwidth,clip]{rv_GJ357_3AP_ALL_02_pcurve_c.ps}

\includegraphics[angle=270, width=0.40\textwidth,clip]{rv_GJ357_OLD_03_pcurve_d.ps}
\includegraphics[angle=270, width=0.40\textwidth,clip]{rv_GJ357_3AP_ALL_03_pcurve_d.ps}
\caption{Estimated posterior probability density based on DRAM samplings of models with $k = 1, 2, 3$ Keplerian signals as a function of the period parameter of the $k$th signal (top to bottom). The red arrow indicates the position of the global maxima in the period space and the horizontal lines denote the 10\% (dotted), 1\% (dashed), and 0.1\% (solid) equiprobability thresholds with respect to the maxima. The left (right) hand side panels denote the posteriors given the old (new) data (see text).}\label{fig:signal_search}
\end{figure*}

Three clear periodic signals were identified in the combined RV data as reasonably unique solutions in the period space between 0.5 days and combined data baseline of 5500 days (Fig. \ref{fig:signal_search}). These signals satisfy our signal detection criteria \citep{tuomi14b} in the sense that they correspond to significant improvements to the model and have parameters, namely amplitude $K$ and period $P$, that are well-constrained from above and below \citep[see e.g.][]{tuomi11b,tuomi14b}. We illustrate the uniqueness of these signals in Fig. \ref{fig:signal_search} and describe their significances in Table \ref{tab:RV_significance}.

\begin{table}
\caption{Logarithms of maximum likelihood values $\log L(k)$ and estimated model probabilities $P(k)$ given all the data assuming equal \emph{a priori} probabilities for models with $k$ Keplerian signals. The table contains model comparisons for the old data sets of \citep[][]{tuomi2019} as well as for analyses where the new UVES data reduction is used together with division of HARPS data into independent time-series for different wavelength intervals (new data). The last column denotes the (potential) period $P_{\rm k}$ of the $k$th signal.}\label{tab:RV_significance}
\begin{center}
\begin{tabular}{lllllllll}
\hline \hline
    & Old data & & New data \\
$k$ & $\log L(k)$ & $P(k)$ & $\log L(k)$ & $P(k)$ & $P_{\rm k}$ (days) \\ 
\hline
0 & -425.43 & 3.1$\times 10^{-6}$ & -531.39 & 1.2$\times 10^{-19}$ & -- \\
1 & -403.13 & 0.051 & -500.90 & 3.7$\times 10^{-12}$ & 9.1 \\
2 & -388.78 & 0.293 & -478.98 & 2.1$\times 10^{-8}$ & 3.9 \\
3 & -375.45 & 0.612 & -451.58 & 0.029 & 55.7 \\
4 & -365.48 & 0.045 & -434.79 & 0.971 & 34.0 \\
\hline \hline
\end{tabular}
\end{center}
\end{table}

As can be seen in Table \ref{tab:RV_significance}, although the signals correspond to unique maxima in the period space (Fig. \ref{fig:signal_search}), only one signal is detected robustly in the old data such that there is \emph{strong evidence} for it according to the requirement of \citet{kass95} that the model is 150 times more probable than its rival. Yet, if we assume circular orbits, all three signals are detected for the old data significantly as well. We also demonstrate in Table \ref{tab:RV_significance} and Fig. \ref{fig:signal_search} that three signals are detected credibly in the new data set. As can be seen, it is clear that when we treat the velocities calculated for independent subsets of orders as independent data sets, this greatly enhances the significance (Table \ref{tab:RV_significance}) of the signals, for instance revealing the presence of the 55~d signal after modelling only the first Keplerian signal (see Fig. \ref{fig:signal_search}). We also note that all three signals are supported by i) all instruments and ii) RVs calculated for both wavelength ranges of the HARPS instrument. This implies that the signals are i) independent of instrument and ii) wavelength-invariant, demonstrating that they are unlikely to have been caused by biases in instrumentation or colour-dependent stellar variability associated with stellar activity \cite[see][]{tuomi2018}. We also note that the potential existence of a fourth signal at a period of 34 days (see Table \ref{tab:RV_significance}) should be investigated when more data becomes available.

We have presented evidence for three unique and significant signals in the combined radial velocity data of GJ~357 at periods of 9.12462$\pm$0.00127, 3.93055$\pm$0.00025 and 55.698$\pm$0.045 days, where the uncertainties correspond to the standard errors. The Keplerian parameters and the inferred minimum masses and semi-major axes are tabulated in Table \ref{tab:RV_parameters} based on the new data set.

\begin{table*}
\caption{Maximum \emph{a posteriori} estimates, standard errors and 99\% Bayesian credibility intervals of the Keplerian parameters and candidate masses given a model with three Keplerian signals and the new data set. The parameters are the Keplerian amplitude $K$, period $P$, eccentricity $e$, the longitude of pericentre $\omega$, mean anomaly $M_{0}$, minimum mass $m \sin i$, and semi-major axis $a$. The uncertainties of minimum masses and semi-major axes have been estimated by accounting for the uncertainty in the stellar mass.}\label{tab:RV_parameters}
\begin{center}
\begin{tabular}{lllllllll}
\hline \hline
& GJ~357 b & GJ~357 c & GJ~357 d \\
\hline
$K$ (ms$^{-1}$) & 2.20$\pm$0.25 [1.54, 2.95] & 1.84$\pm$0.23 [1.18, 2.51] & 2.41$\pm$0.32 [1.52, 3.31] \\
$P$ (days) & 9.1246$\pm$0.0013 [9.1207, 9.1279] & 3.93055$\pm$0.00025 [3.92981, 3.93129] & 55.698$\pm$0.045 [55.555, 55.826] \\
$e$ & 0.072$\pm$0.053 [0, 0.240] & 0.047$\pm$0.059 [0, 0.267] & 0.033$\pm$0.057 [0, 0.259] \\
$\omega$ (rad) & 5.3$\pm$1.9 [0, 2$\pi$] & 3.9$\pm$1.3 [0, 2$\pi$] & 0.3$\pm$2.2 [0, 2$\pi$] \\
$M_{0}$ (rad) & 4.3$\pm$1.8 [0, 2$\pi$] & 5.8$\pm$2.2 [0, 2$\pi$] & 5.3$\pm$2.0 [0, 2$\pi$] \\
\hline
$m \sin i$ (M$_{\oplus}$) & 3.68$\pm$0.48 [2.35, 5.14] & 2.32$\pm$0.33 [1.40, 3.24] & 7.20$\pm$1.07 [4.39, 10.33] \\
$a$ (AU) & 0.0607$\pm$0.0021 [0.0544, 0.0664] & 0.0348$\pm$0.0012 [0.0310, 0.0379] & 0.2040$\pm$0.0069 [0.1820, 0.2218] \\
\hline \hline
\end{tabular}
\end{center}
 \label{tab:system}
\end{table*}

In order to help validate our results, we performed an independent test of the reality of the system by running the Exoplanet Mcmc Parallel tEmpering Radial velOcity fitteR  \texttt{EMPEROR}\footnote{https://github.com/ReddTea/astroEMPEROR} \citep{pena19} automatic signal detection code.  \texttt{EMPEROR} employs a similar statistical model to the DRAM approach, except the exponential moving average smoothing time-scale is left as a free parameter, and in this case we also used a first-order moving average correlated noise model. The main difference between the two methods is that \texttt{EMPEROR} performs the MCMC samplings by using parallel tempering methods across multiple simultaneous chains (five in this case), each one hotter (with smaller $\beta < 1$ such that likelihood function $l^{\beta}$ is used rather than $l$) than the next. It uses the \texttt{EMCEE} code \citep{foreman13} as the basis for the samplings in this respect. The statistical model is taken from the cold chain ($\beta = 1$). We employed 150 walkers and 15'000 steps per chain, meaning we obtain a chain length of 11.25M, along with half that value used as burn-in, allowing \texttt{EMPEROR} to thoroughly explore the high dimensional posterior parameter space.  \texttt{EMPEROR} yields a similar solution to the DRAM samplings for the outer two planet candidates, arriving at the 55~day planet signal first, followed by the 9~day planet signal next. The transiting 3~day planet signal was not detected in the automatic runs, (by automatic runs we mean the basic \texttt{EMPEROR} operational format where the code runs alone by modifying the priors for each model test, testing the Bayesian significancies, for example, without any human intervention), meaning we confirm the existence of the two outer signals, but could not confirm the RV signal from the transiting planet.

\section{Stellar Activity Analysis}\label{act}

\subsection{Spectroscopic Activity indices}
In order to test the true planetary nature of the signals found in the RVs, we analysed the spectral activity indices derived from the available spectra. First, for the HARPS data we computed the activity indices using the HARPS-TERRA algorithm \citep{AngladaButler2012}. The full-width at half maximum (FWHM) of the cross-correlation function (CCF) and the bisector inverse slope (BIS) are read directly from the fits headers obtained as a product of the HARPS online data reduction software (DRS). The rest of the activity indices, i.e. S-index, H$_{\alpha}$, Na I, Na II, are not provided as a standard product of the DRS and thus are computed from the one dimensional reduced spectra. For instance, the S-index values that are calculated by studying the cores of the Calcium~\sc ii\rm ~H \& K lines (e.g. \citealp{jenkins08,jenkins11a,sousa11}), are calibrated to the Mount Wilson system for direct comparison to other stars, giving rise to a proxy for the chromospheric activity variability of the star. 

We ran a Generalized Lomb-Scargle (GLS; \citealp{zechmeister09b}) periodogram on each of the activity indices time series. For each periodogram we defined a pseudo-Nyquist frequency $\omega_{\rm max}=\pi/\Delta T$, i.e., the largest frequency this analysis is sensitive to. $\Delta T$ is the median spacing between data points in the time series. We also defined our optimal minimum frequency by $\omega_{\rm min}=2\pi/T_{\rm max}$, with $T_{\rm max}$ the time span of the time series. We defined our grid of frequencies by setting the spacing as $\Delta \omega=\eta\, \omega_{\rm min}$, with $\eta=0.1$. Finally the number of sample frequencies to test is $n_{\omega}=(\omega_{\rm max} - \omega_{\rm min})/\Delta \omega$.

\begin{figure}
    \centering
    \includegraphics[width=\columnwidth,clip]{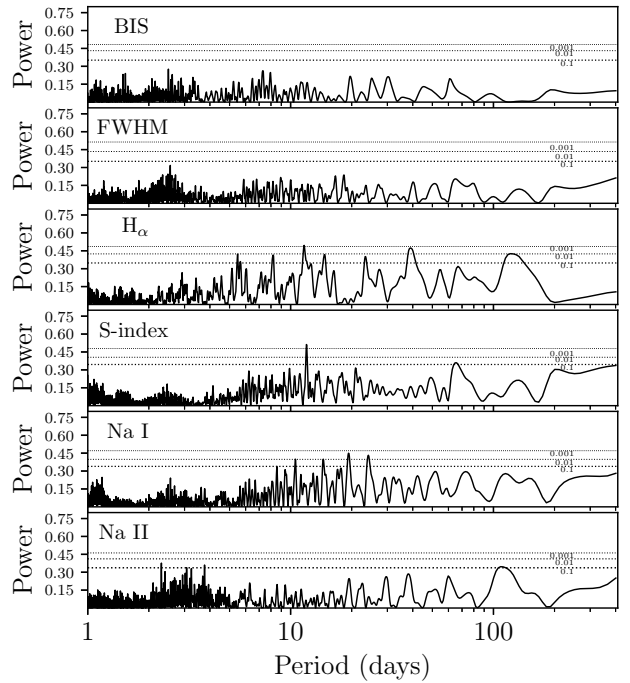}
    \caption{{\it Top to bottom}: Generalized Lomb-Scargle periodogram for the BIS, CCF-FWHM, H$_{\alpha}$, S-index, Na I and Na II derived from the spectra using the HARPS-TERRA reduction package. Dotted lines, from bottom to top, represent the  10, 1 and 0.1\% significance levels computed by running 5000 bootstrap iterations.}
    \label{fig:activity_pg}
\end{figure}

\begin{figure}
    \centering
    \includegraphics[width=\columnwidth,clip]{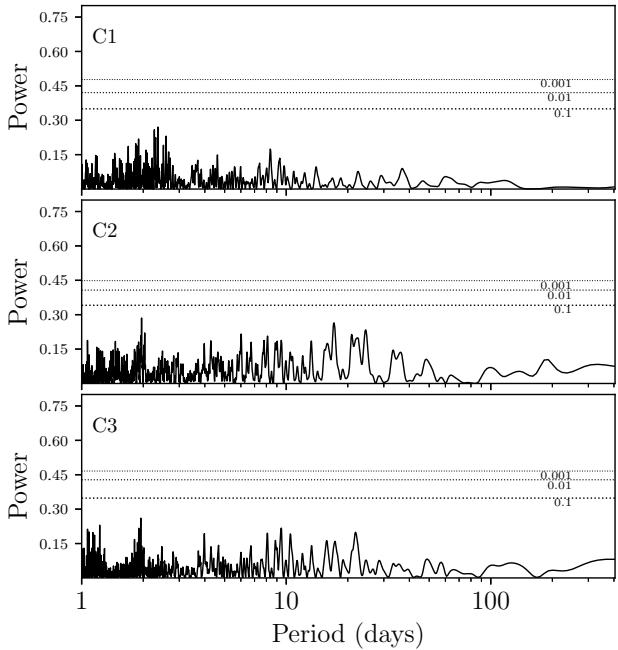}
    \caption{{\it Top to bottom}: Generalized Lomb-Scargle periodogram of the colour dependent differential velocities derived from the HARPS spectra using the HARPS-TERRA software. Dotted lines, from bottom to top, represent the  10, 1 and 0.1\% significance levels computed by running 5000 bootstrap iterations.}
    \label{fig:diff_vels_pg}
\end{figure}

Figure \ref{fig:activity_pg} shows the results of the GLS periodogram for the activity indices described before. No statistically significant power peaks are found in the power spectrum of BIS, FWHM, and Na II.  We note that the reason for the lack of a signal in the BIS and FWHM indices may be due to the general lack of any well constrained continuum region for HARPS M dwarf CCFs, a noise source that could mask any signal present (e.g. see Fig.~7 in \citealp{berdinas17}).   From the periodogram of H$_{\alpha}$ significant powers are found at $\sim$75 and $\sim$115 days. The S-index periodogram shows a significant power above the 0.1\% significance level at $\sim$73 days. Na I, on the other hand, exhibits power peaks above the 1\% significance level at $\sim$120 and $\sim$150 days.  Similar values were reported by \citet{defru17} ($P_{rot}$ = 144~d) and \citet{schofer19} ($P_{rot}$ = 74.3~d).  We believe that the S-indices here are probing the primary rotation period of the star, with a periodicity in good agreement with that presented in \citep{luque19} using both photometric and spectroscopic activity measurements.  Beyond these standard indices we also perform the same analysis on our new colour dependent differential velocity indices (Fig.~\ref{fig:diff_vels_pg}), revealing no significant periods. 

\subsection{ASAS Photometry}
\begin{figure}
\center
\includegraphics[scale=0.5,clip]{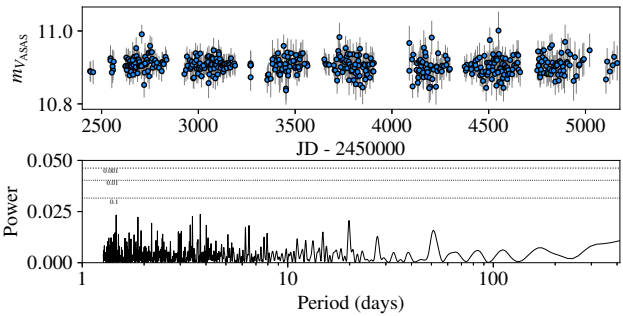}
\caption{{\it Top}: $V$-band ASAS photometry for GJ~357. {\it Bottom}: Generalized Lomb-Scargle periodogram of the photometry. Dotted lines correspond, from bottom to top, to the 10, 1 and 0.1\% significance levels computed via 5000 bootstrap iterations on the original photometry time series.}
\label{fig:ASAS_pg}
\end{figure}

We searched for photometric measurements of the star from the All-Sky Automated Survey \citep{pojmanski97} catalog. We found 644 measurements consisting of $V$-band photometry spanning $\sim$9 years, from Nov 20th 2000 to Nov 29th 2009. We filter the photometry by selecting the highest quality data, flagged as ``A'' and ``B''. After filtering out 54 bad data points, we analyse 587 photometric measurements to search for stellar rotational periods. 
Figure \ref{fig:ASAS_pg} shows the GLS periodogram of the photometry showing no statistical significant power peaks to constrain $P_{\rm rot}$ for this star.

\section{TESS Confirmation}\label{tess}

The TESS is a NASA-sponsored Astrophysics Explorer-class mission that is performing a wide-field survey to search for planets transiting bright stars \citep{TESS}.  It has four $24\times24\degr$ field of view cameras with four 2k$\times$2k CCDs each, with a pixel scale of 21 arcseconds per pixel and a bandpass of 600-1000\,nm. 
GJ~357 was observed by TESS in Sector~8 using CCD~3 of Camera~2 between February 2nd and 27th 2019.

\begin{figure}
    \centering
    \includegraphics[width=\columnwidth]{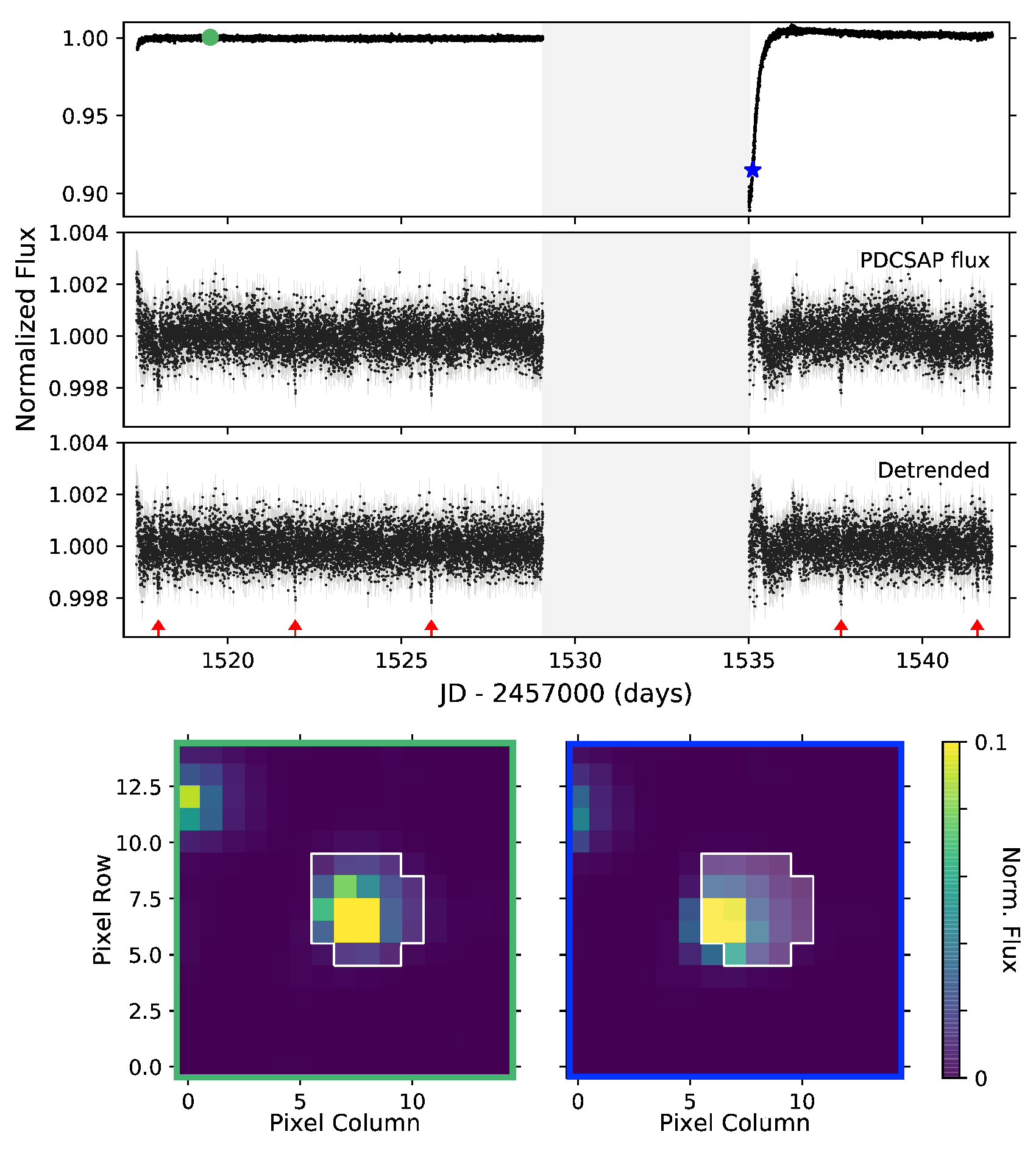}
    \caption{{\it Top}: GJ~357 TESS light curve obtained with aperture photometry from calibrated pixels given in the Target Pixel files. The grey area highlights $\sim$6~days of no observations (see text for details). {\it Second}: Normalised PDC photometry without 4-sigma outliers. {\it Third}: Final PDC light curve after accounting for the long-term variability. The red arrows indicate the transit times of GJ~357~c. {\it Bottom}: Target Pixel files at epochs 2458519.49 (left; green frame) and 2458535.11~days (right; blue frame) corresponding to the green dot and a blue star in the top panel. The shaded white areas indicate the aperture used to the get the light curve in the top panel.}
    \label{fig:tess}
\end{figure}

The \textit{TESS} data of GJ~357 were obtained from the NASA Mikulski Archive for Space Telescopes (MAST)\footnote{\url{http://archive.stsci.edu/tess/bulk_downloads/bulk_downloads_ffi-tp-lc-dv.html}.}. We downloaded the corresponding Target Pixel (TP) file provided for our target. This TP file contains raw and calibrated fluxes in an image cube of $15\times15$ pixels observed with a two-minutes temporal cadence. After removing cadences affected by instrumental issues and marked with non-zero ``quality'' flags, the temporal dimension of the cube comprised 13392 epochs. The top panel of Fig.~\ref{fig:tess} shows the light curve obtained by performing simple aperture photometry. The aperture was defined by the pixels with flux above the 85~th percentile of the median image calculated along the time axis.

The observations of GJ~357 were paused during the spacecraft perigee passage to download the data. Observations should have been resumed soon after, but an interruption in communications between the instrument and spacecraft caused a final $\sim$6~days gap in the light curve (JD=241529.06--241535.00). Also, the need of turning on the camera heaters after the communications failure caused a small drift in the star position due to changes in the camera focal plane (see the star drift at the bottom panels of Fig.~\ref{fig:tess}). As a consequence, the light curve shows a progressive increase of the flux after resuming the observations. The flux returned to nominal within a few days. 

We tried different apertures to mimic this effect as well as a simple spline fitting, but finally we decided to use the Presearch Data Conditioning (PDC) flux given by the Science Processing Operations Center \citep[SPOC,][]{jenkinsjm16} at MAST public archive. The PDC pipeline was first implemented for \emph{Kepler} to account for systematics error sources from either the telescope or the spacecraft, such as focus changes, sudden pixel sensitivity dropouts, pointing drifts, and thermal transients. The second panel of Fig.~\ref{fig:tess} shows the normalised PDC light curve resulting after discarding high-dispersion epochs with a 4-sigma criteria. Besides the transit features of the inner planet (marked with red arrows), the PDC light curve shows evident long term variabilities. We smoothed the light curve with a Savitzky-Golay filter \citep{savitzkygolay}. The third panel of Fig.~\ref{fig:tess} shows the final detrended light curve.  Although there are three stars within TESS pixel size of 21$''$, our target star and two other Gaia sources, these background contaminants are around 7~magnitudes fainter in the optical, rendering their contamination factor small. In \citet{luque19} they also considered the effect these background stars would have on the TESS lightcurves, in order to determine the dilution factor needed to correct for this effect and calculate the proper planetary radius.  They found a value consistent with zero, therefore no dilution correction is necessary.

\begin{figure}
	\includegraphics[angle=90, width=\columnwidth]{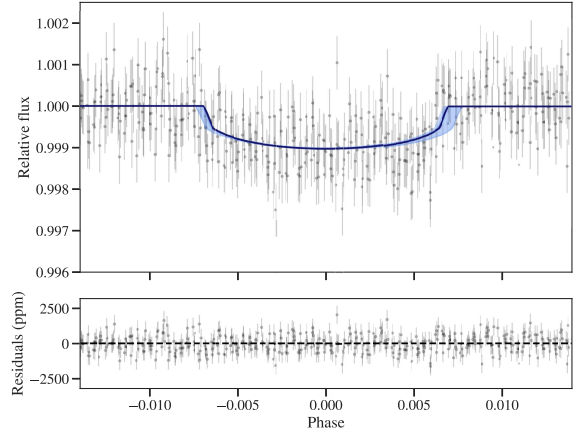}
    \caption{Phase folded TESS photometry. The solid blue line is the best fit for the photometry and the blue shaded regions represent the $1,2,3\,\sigma$ confidence levels. Bottom: The residuals of the fit in ppm.}
    \label{fig:TESSLC}
\end{figure}

This light curve was used to perform a joint photometry and RV model fit using \texttt{Juliet} \citep{juliet}. \texttt{Juliet} is a Python tool designed to analyse transits, RVs or both simultaneously, allowing for multiple photometry and RV instruments to be modelled at the same time. It calculates the Bayesian Evidence of the model with the use of Nested Sampling, Importance Nested Sampling, or Dynamic Nested Sampling algorithms \citep{MultiNest, PyMultiNest}, sampling the posterior parameter space as a by-product. For the light curve model, \texttt{Juliet} employs \texttt{batman} \citep{batman}. The Keplerian model is provided by \texttt{RadVel} \citep{radvel}. \texttt{Juliet} also allows the use of Gaussian Processes to model the photometry and RVs. In this work we used a Matern multiplied by exponential kernel from \texttt{celerite} to fit the photometry \citep{celerite}. We considered two different models for the joint photometry and RV fit: one with free eccentricity and another with the eccentricity fixed to 0; adopting the same thresholds for weak and strong evidence as in \cite{juliet} ($\Delta \ln Z = 2$ and $\Delta \ln Z = 5$ respectively).  We found that $\Delta \ln Z = 9.676$ in favour of the fixed eccentricity model and thus we selected it as our final solution. We show the TESS light curve transit event in Figure \ref{fig:TESSLC} and the resulting parameters for GJ~357~$c$ in Table \ref{tab:planet}.

\begin{table}
	\centering
	\caption{Planetary Properties for GJ~357~$c$}
	\begin{tabular}{lc} % six columns, alignment for each
	Property	&	Value \\
	\hline
%    \multicolumn{3}{l}{Light curve Parameters}\\
    P (days)		&	$3.93086\pm0.00004\,[3.93050,3.93101]$  \\\vspace{2pt}
	T$_C$ (BJD - 2450000)&	$8517.9994\pm0.0001\,[8517.9985,8518.0004]$	\\\vspace{1pt}
    $a/R_{*}$		& $20.165^{+0.155}_{-0.190}\,[17.707,20.643]$ \\\vspace{2pt}
    $b$ & $0.529^{+0.010}_{-0.005}\,[0.514,0.600]$    \\\vspace{2pt}
%    \multicolumn{3}{l}{RV parameters}\\
	K (\ms) 	& $1.7372^{+0.0054}_{-0.0007}\,[1.7356,1.7539]$	\\
    e 			& $0$ {\it (fixed)} \\\vspace{2pt}
%    \multicolumn{3}{l}{Planetary Parameters}\\
    M$_p$ (\me)& $2.248^{+0.117}_{-0.120}\,[1.930,2.547]$	\\\vspace{2pt}
    R$_p$ (\re)& $1.167^{+0.037}_{-0.036}\,[1.073,1.264]$ \\\vspace{2pt}
    $\rho_{p}$ (\gccc) & $7.757^{+0.889}_{-0.789}\,[5.867,10.295]$ \\\vspace{2pt}
    a (AU) & $0.033\pm0.001\,[0.028,0.036]$ \\\vspace{2pt}
    $i$ (deg)  & $88.496^{+0.025}_{-0.043}\,[88.063,88.561]$ \\ \hline
%    \multicolumn{3}{l}{Table note here}\\
	\end{tabular}
    \label{tab:planet}
    Values inside the square brackets represent the 99\% posterior estimates from the joint modeling.
\end{table}

\section{Orbital dynamics}\label{dynamic}

With the RV data and photometric TESS observations it is possible to perform a first full characterisation that includes planetary dynamics and orbital analysis. In particular, we sought to constraint the mass of the components and their mutual inclinations. In addition, we studied the global stability of the system in order to provide the most likely dynamical configuration.

\subsection{The impact parameters of planets GJ~357b and GJ~357d}
\begin{figure}
	\includegraphics[angle=0,width=\columnwidth]{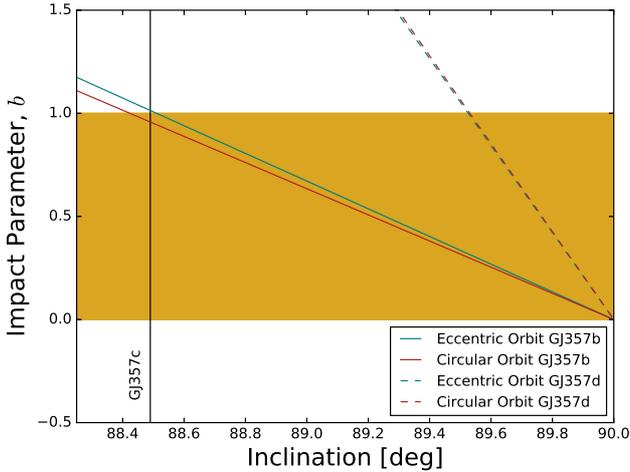}
    \caption{Impact parameters of planets GJ~357~b and GJ~357~d for different inclinations according to equation \ref{eq:1}. Both eccentric (blue line) and circular
    (red line) orbits are shown, given from the values determined by the radial-velocity solution. The yellow-shaded region delimits the solutions allowed 
    for each planet to transit its host star, and the vertical line marks the inclination of the transiting planet GJ~357~c.}
    \label{fig:impact}
\end{figure}

Since only the innermost planet GJ~357c is transiting, we wonder about the  
inclinations of the outermost planets GJ~357b and GJ~357d. In transiting exoplanets, the impact parameter, $b$, 
is the sky-projected distance conjunction, in units of stellar radius \cite[e.g.,][]{josua2010}:

\begin{equation} \label{eq:1}
 b=\frac{a\cos{i}}{R_{\star}}\left(\frac{1-e^{2}}{1+e\sin{\omega}}\right) ,  
\end{equation}
where $a$ is the semimajor axis, $i$ is the orbital inclination, $R_{\star}$ is the radius of the star, $e$ is the eccentricity and $\omega$ is 
the longitude of pericenter. The $b$-parameter varies from $b=0$, when the planet cross the centre of the stellar disk, and $b=1$, when 
it is on the cusp of the disk. We analysed the $b$-parameter for the two outermost planets to search for the maximum inclinations that make these planets non-transiting.
For each planet, we analysed both circular and eccentric orbits, where the chosen eccentric orbits were the nominal values provided in Table \ref{tab:system}, i.e., $e_{b}$=0.072 and $e_{d}$=0.033. 
We find that planet GJ~357~b has a maximum inclination ranging between $88.5\degr-88.4\degr$.
For planet GJ~357~d, we found that the maximum angle is $\sim89.5\degr$. Since the inclination found by the analysis of the TESS lightcurve for the inner planet GJ~357~c 
is $i\sim88.49\degr$, we note that even a near-coplanar configuration makes the two outer planets non-transiting (see Fig. \ref{fig:impact}). This may explain the lack of transit signals 
for planets GJ~357~b and GJ~357~d in TESS data. 

In fact, the general belief is that planets tend to reside in nearly coplanar orbits within a few degrees of inclination from an inertial
plane perpendicular to the total angular momentum of the system. For example, in the Solar System, the highest inclination is obtained for Mercury, which has $i\sim7\degr$.  
This hypothesis is also supported observationally. Indeed, the combination of \emph{Kepler} and HARPS data suggest that the mutual inclinations of multi-planet systems are $\lesssim$5$\degr$ \cite[see e.g.,][]{fabrycky2014,tremaine2012,figueira2012}. However, while stellar and planetary formation theories require an accretion disk 
perpendicular to the total angular momentum of the system in which planets form, close encounters and/or captures in mean motion resonances (MMRs) orbits
might significantly increase their eccentricities and inclinations \cite[e.g.,][]{lee2007,libertA2009}. With this in mind, we conducted 
dynamical simulations to explore the minimum values of the inclinations for planets GJ~357~b and GJ~357~d. Furthermore, their masses 
are dependent on their orbital inclination angles, which means that any uncertainties in the inclination angles will seriously affect their determined masses. Hence, exploring 
the minimum values of their inclination angles is equivalent to exploring the maximum values of their masses. 

\subsection{The MEGNO criterion and short-term stability}

\begin{figure}
	\includegraphics[angle=0,width=\columnwidth]{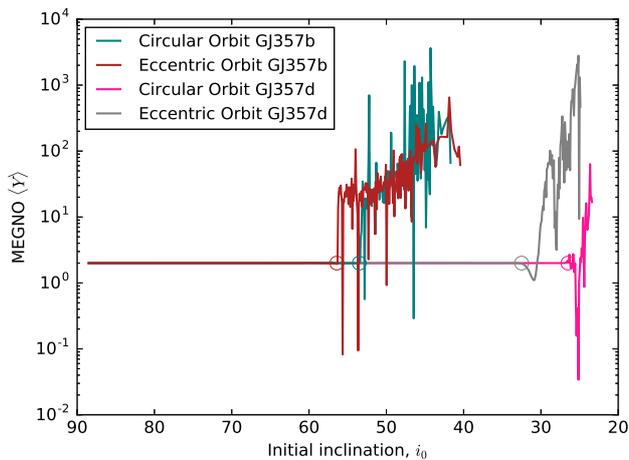}
    \caption{Stability analysis based on the MEGNO chaos index for different inclinations of planets GJ~357~b and GJ~357~d.
    Both eccentric and circular orbits are explored for 1000 initial inclination angles ranging from 88.5$\degr$ to 0.1$\degr$.
    The systems are stable while $\langle Y(t) \rangle \rightarrow 2$, and become unstable as $\langle Y(t) \rangle$ diverge from 2.
    The circles over the horizontal lines give the angles when the divergence start, i.e., when the system is no longer stable.}
    \label{fig:megno-inc}
\end{figure}

We study the short-term stability of the system 
through the Mean Exponential Growth factor of Nearby Orbits, $Y(t)$ (MEGNO, \cite{cincottasimo1999,cincottasimo2000,cincotta2003}).
MEGNO is a chaos index that has been extensively used within dynamical astronomy, in both the Solar System and extrasolar planetary systems \cite[e.g.,][]{jenkins09,hinse2010,contro2016,gunter2019}. It allows us to explore the whole parameter 
space at small computational cost. In brief, MEGNO evaluates the stability of a body's
trajectory after a small perturbation of the initial conditions. Its time-averaged mean value,
$\langle Y(t) \rangle$, amplifies any stochastic behaviour, allowing the detection of hyperbolic 
regions during the integration time. Therefore, $\langle Y(t) \rangle$ allows us to distinguish between chaotic and quasi-periodic trajectories: if $\langle Y(t) \rangle \rightarrow \infty$ for $t\rightarrow \infty$ the system is chaotic; while if $\langle Y(t) \rangle \rightarrow 2$ for $t\rightarrow \infty$ the motion is quasi-periodic. We used the MEGNO implementation within the N-body integrator {\scshape rebound} \citep{rein2012}, which made use of the Wisdom-Holman WHfast code \citep{rein2015}. We evaluated the inclination 1000 times between 88.5$\degr$ and 0.1$\degr$ for both planets by considering circular and eccentric orbits (see Fig.~\ref{fig:megno-inc}).
The integration time was set to $10^{6}$ times the orbital period of the outermost planet, GJ~357~d. The time-step was set to $5\%$ of the 
orbital period of the innermost planet, GJ~357~c. In order to prevent unnecessary computations, we prematurely terminated any integration when 
one of the planets was ejected from the system, where we consider an ejection cases when the semimajor axis of a given planet reaches values larger than 20~AU, hinting at clear chaotic behaviour. 

Using the method outline, we firstly evaluated the inclination 
of the planet GJ~357~b by assuming the (fixed) planetary parameters given in Tables 
\ref{tab:system} and \ref{tab:planet}, and considered co-planar orbits between GJ~357~c and GJ~357~d.  We find that 
planet GJ~357~b becomes unstable when the inclination is less than or equal to $\sim$56$\degr$ for eccentric orbits with $e_{b}=$0.072, and 
$\sim$53$\degr$ or less for circular orbits. Secondly, we explored the 
outermost planet GJ~357~d, and held constant the values of planet GJ~357~c and GJ~357~b (given in Tables \ref{tab:system} and \ref{tab:planet}), considering coplanar orbits. 
Since planet GJ~357~d is very far from the two innermost planets, it is expected to withstand lower inclinations. Indeed, we find that for eccentric orbits 
of $e_{d}=$0.033, the minimum inclination is $\sim$32$\degr$, and for circular orbits, it is $\sim$26$\degr$.

\begin{figure}
	\includegraphics[angle=0,width=\columnwidth]{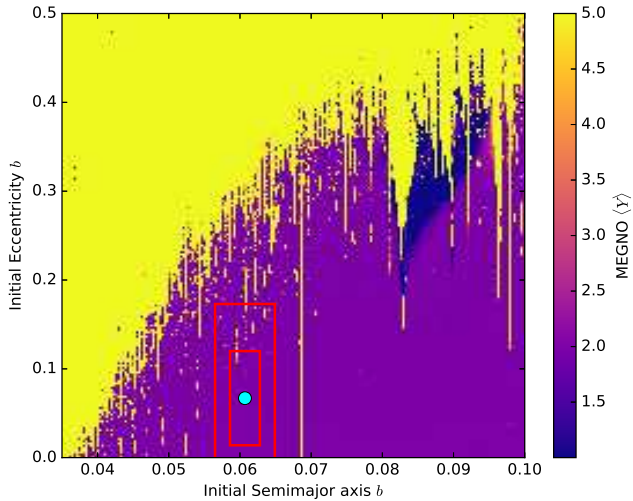}
    \caption{Dynamical analysis of planet GJ~357~b based on a MEGNO-map for a coplanar configuration.
    The size of the map is 200$\times$200 pixels, which explores the $e_{b}-a_{b}$ parameter space.
    When $\langle Y(t) \rangle \rightarrow 2$ (purple shaded regions) quasi-periodic orbits are found, and
    chaotic systems are found when $\langle Y(t) \rangle \rightarrow 5$ (yellow shaded regions). The nominal value for GJ~357~b is shown by the
    blue marker, while the 1$\sigma$ and 2$\sigma$ uncertainties from the radial-velocity orbital solution are shown by the red boxes.}
    \label{fig:megno-map}
\end{figure}

From these sets of simulations, we note that in the nearly co-planar scenario, the stability of the system is mainly controlled by the two innermost planets GJ~357~c and GJ~357~b, while the outermost
planet GJ~357~d is too distant to induce perturbations. However, while planet GJ~357~c is strongly characterised by the synergy of both the RV and photometric measurements, planet GJ~357~b is only characterised through its RV. Hence, we wonder if in the range of uncertainties obtained for planet GJ~357~b, the system could actually be fully stable. To address this 
question we constructed a two-dimensional MEGNO-map in the $e_{b}-a_{b}$ parameter space. The integration time and the time-step are the same
as used before. In order to reduce the computation time, we checked the integration every 1000 yr, and we chose to stop a given 
integration when $\langle Y(t) \rangle \textgreater 5$ \citep{hinse2015}. The size of the obtained MEGNO-map was 200$\times$200 pixels, meaning 
we explored the $e_{b}-a_{b}$ parameter space for planet GJ~357~b up to 40,000 times (Fig.~\ref{fig:megno-map}). We find that planet GJ~357~b seems to be 
fully stable in the 1$\sigma$ uncertainty space (inner red-box), but some instabilities might be found 
when we extend to the 2$\sigma$ uncertainty space (outer red-box), which mostly corresponds to larger values of 
$e_{b}$. This result suggests that in terms of stability, planet GJ~357~b favours  
low eccentricity values in terms of the radial velocity orbital solution. 

However, an alternative configuration to the nearly coplanar scenario is still possible. Indeed, planets GJ~357~b and GJ~357~c are close 
to a MMR of 7:3, i.e. a high-order resonance of $q$=4. \cite{libert2009} suggested that high-order 
resonances might provoke an excitation of inclinations angles, where the maximum mutual inclinations range between 20$\degr$ and 70$\degr$. This may explain 
why planet GJ~357~b is not transiting: the two inner planets are trapped in an inclination-type resonance. To check if 
this is the case, we computed the resonance angles defined as $\phi_{i}=(p+q)\lambda_{2}-p\lambda_{1}-q\omega{_i}$, where $\lambda_{i},~i=1,2$ is 
the mean longitude of each planet and $\omega_{i}, i=1,2$ are their longitudes of perihelion, following \cite{millho2018}. The behaviour of these angles can show whether a 
system is in resonance or not: if at least one of the angles oscillates then the system is inside the resonance. With this aim we conducted 
N-body integrations for $10^5$ years, which is equivalent to $\sim$4 million orbits 
of planet GJ~357~b. We used the nominal values given in Table \ref{tab:system} and \ref{tab:planet}, and a set of 10 initial inclinations for planet b ranging from 
$88.5\degr$ to $56.0\degr$. For all of the scenarios tested, we found non-librating behaviour of the resonance angles, 
indicating that the system is not trapped in a 7:3 resonance. 

\subsection{Long-term stability}

The set of results obtained so far allowed us to explore the long-term stability of the system in 
a smaller parameter space. Again we used the MEGNO criterion to find the most stable solutions, and we performed simulations up to 5$\times10^{7}$ yr, storing the value of MEGNO every 1000 yr. We determined that any given system showed a trend towards chaotic behaviour 
when $|\Delta \langle Y(t) \rangle|> 0.5$, where $\Delta \langle Y(t) \rangle =2.0-\langle Y(t) \rangle$, i.e., when the integration started to diverge from values of 2.0, particularly beyond 0.5 units. Next, we considered oscillations of $\langle Y(t) \rangle$ around 2.0 with amplitudes of $<$ 0.5 as stable systems 
that need more integration time to clearly reveal their real trends. Two reasons prevented us from carrying out longer simulations of hundreds or thousands of Myr; first, the high computational cost, and second, the large number of errors propagated along the integrations will eventually imply that all the simulated scenarios show some degree of instability. Indeed, for very long-term integrations, a given chaotic behaviour might be caused by a lack of energy conservation accumulated at each time-step \citep{rein2015}.

We integrated the system considering co-planar orbits, where the two inner planets had 
circular orbits. We found that the system was stable for the total integration time, with 
a $|\Delta \langle Y \rangle|\sim 0.2$. Moreover, previously we found that planet GJ~357~b should 
have a low eccentricity in the 1$\sigma$ uncertainty region to ensure the short-term stability of the system. Therefore, we explored different eccentricities: 0.072, 0.06, 0.05, 0.04, 0.02, 0.01, 0.005, and 0.001. We found that only very low eccentricities $<$0.01 keeps our criteria of stable solutions. As such, we find that only near-circular configurations for planet GJ~357~b ensure long-term stability. 

We moved on to examine the minimum inclination of GJ~357~b. Previous results regarding 
its short-term stability showed a minimum value of $i_{b}\sim53\degr$ when circular orbits were assumed for the two innermost planets b and c. However, during long-term integrations we found that the system became unstable when approaching $10^{6}$ yr. As such, we explored the minimum values of $i_{b}$ starting from 53$\degr$ and increasing in increments of 2$\degr$ until we found a stable solution. We find that for $i_{b}\sim60\degr$, the system is stable with $|\Delta \langle Y \rangle|\sim 0.3$. This means we can place strong constraints on the maximum mutual inclination of the two inner planets GJ~357~b and GJ~357~c of 28.5$\degr$, that yield a maximum mass for the planet GJ~357~b of 4.25~\me. 

We next explored the non-coplanarity of the outermost planet GJ~357~d. We considered the two inner planets as being coplanar, and residing in circular orbits. During the short-term stability analysis we found the minimum inclination of this planet as $i_{d}\sim26\degr$ for circular orbits, and $i_{d}\sim32\degr$ when $e_{d}=0.033$. During these long-term integrations we found that for a circular configuration, the system becomes unstable in the range of $\sim10^{6}$ yr. Following the same strategy described before, we progressively increased the inclination in increments of 2$\degr$ to investigate its long-term stability.  We find that for $i_{b}\sim40\degr$ or more, the system is stable with $|\Delta \langle Y \rangle|\sim 0.1$, i.e., the maximum mutual inclination between the two innermost planets and GJ~357~d is 48.5$\degr$, which implies a mass equivalent to 11.2~\me\, for this planet. On the other hand, in the case of a non-eccentric orbit for GJ~357~d when $i_{d}=40\degr$, we find the system becomes unstable in the range of 2-3$\times10^{7}$ yr for eccentricities ranging from 0.01 to 0.033. When we tried with higher inclinations up to $i_{d}=50\degr$, we find that the system remains stable up to 4-5$\times10^{7}$ yr, but its rapidly changing behaviour indicates instability at the limit of our integration. Therefore, at this moment, we can not unambiguously determine the maximum inclination in the case of eccentric orbits for GJ~357~d.

The set of results obtained in this section seems to favour a system of two 
super-Earths of $M_{c}=2.087$~\me, $3.69<M_{b}<4.25$~\me\, and a likely third mini-Neptune planet of $7.20<M_{d}<11.20$~\me, where the innermost planets 
GJ~357~c and GJ~357~b reside in near-circular orbits. The analysis of the $b$-parameter, and the normal trend for planetary systems to be nearly-coplanar, encouraged us to favour the hypothesis of near-coplanar orbits, which correspond to the 
minimum values in our mass ranges. However, we also found that the system might remain stable for non-coplanar configurations if the mutual inclinations do not exceed 28.5$\degr$ for planets GJ~357~c and b, and 48.5$\degr$ for planets GJ~357~c and d, which gives rise to the maximum values of the planetary masses.

\begin{figure}
	\includegraphics[angle=0,width=\columnwidth]{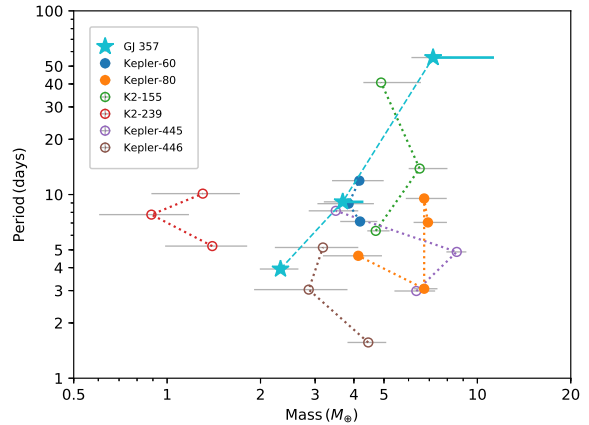}
    \caption{Mass versus orbital period for planetary systems hosting three or more low-mass planets with absolute masses measured to better than 43.5\% (the uncertainty for GJ~357~d).  Dashed lines connect each of the systems, uncertainties are shown as light grey error bars, and the thick and coloured lines mark the mass upper limits for GJ~357~b and d.  Only GJ~357 has planets with absolute masses detected by radial-velocities (star symbols), the others have either been detected by transit photometry or transit timing variations (circular symbols).  Open symbols mark those systems where the mass estimates were drawn from mass$\--$radius relationships.  The key in the plots highlights each of the planetary systems (K2-155$\--$\citealp{diez18a}; K2-239$\--$\citealp{diez18b}; Kepler-60$\--$\citealp{steffen13,jontof16}; Kepler-80$\--$\citealp{macdonald16}; Kepler-445 \& Kepler-446$\--$\citealp{muirhead15,mann17}).}
    \label{fig:mass-period}
\end{figure}

With these strong upper mass constraints provided by the dynamic analysis, the GJ~357 planetary system becomes the only such multi-planet system that contains at least three planets that has been detected by RVs at this level of precision, that we are currently aware of (see Fig.~\ref{fig:mass-period}).  Transit photometry and transit timing variations have discovered the remaining systems.  It can be seen in this figure that GJ~357 provides one of the most well constrained system of planets in mass and period parameter space.  This is largely due to the brightness of the target that has allowed a vast quantity of precision RVs to be observed, along with the detailed modelling effort that we provide.  

When we compare the GJ~357 system to the other multi-planet systems shown in Fig.~\ref{fig:mass-period}, we can see that the configuration is generally not typical of the population.  Each of the planets gets more massive as a function of their orbital period.  This likely reflects the bias inherent in the RV method, since most other systems show fairly vertical relationships, meaning the planet masses in these other systems are broadly similar, regardless of orbital period; there is no trend.  In fact, Kepler-445 is the only system here where the outer planet is the least massive one, statistically.  We can also see that there exists only one multi-planet system with these constraints with masses below 2~\me, K2-239.  Actually, GJ~357c is the lowest mass planet discovered in such multi-planet systems, outside of K2-239.  The plot also highlights that GJ~357 is only the third star to host such a rich and well constrained planetary system, (after Kepler-60 and Kepler-80), where masses were not drawn from mass$\--$radius relationships.

\section{Conclusions}\label{sum}

The bright and nearby M dwarf star GJ~357 has recently been shown to host a low-mass transiting planetary system using precision photometry from the TESS mission and radial-velocity data.  Our analysis of new precision radial-velocities from three different instruments, the HARPS, the HiRES, and the UVES results in evidence for a system of three small planets orbiting the star with periods of approximately 9, 4, and 56 days.  Therefore, we confirm the detection of the planetary system in radial-velocity data independently of the TESS photometry.  The detected planets have minimum masses of 3.68$\pm$0.48, 2.32$\pm$0.33, and 7.20$\pm$1.07~\me\, for the planets GJ~357b,c,and d, respectively.  When we simultaneously fit the TESS lightcurve with the radial-velocities, we find tight constraints on the mass, radius, and density of $2.248^{+0.117}_{-0.120}$~\me, $1.167^{+0.037}_{-0.036}$~\re, and $7.757^{+0.889}_{-0.789}$~\gccc\, respectively, for planet GJ~357c.  Comparison to structure models indicates the planet is a predominantly iron-rich world.

In order to better understanding the real nature of this system and unveil its most likely 
dynamical configuration, we combined both short- and long-term simulations to constrain the planetary masses, eccentricities, and inclinations in terms of their stability based on the MEGNO criterion. The simulations revealed the following:

\begin{enumerate}

\item The system is likely coplanar, or near-coplanar, since the analysis of the impact parameter revealed that closely coplanar orbits would render the two outer planets such that they would not be expected to transit the star.
\newline

\item The two innermost planets should be in near-circular orbits. Indeed, we found that to ensure the long-term stability of the system the eccentricity of planet GJ~357b should be $\lesssim$0.01.
\newline

\item The two non-transiting planets, GJ~357b and d, can not have inclinations exceeding 60$\degr$ and 40$\degr$, respectively. This allows upper limits to be placed on the masses of these planets of 4.25~\me and 11.2~\me, respectively.
\newline

\end{enumerate}

The GJ~357 planetary system adds to the growing number of systems orbiting some of the smallest and nearest stars. The system we have uncovered appears to be composed of a mix of rocky planets and a mini-Neptune.  This is the only such multi-planet system with well constrained masses detected by RVs, and only the second system that does not apply empirical or theoretical mass$\--$radius relationships to calculate the planetary masses.  Futhermore, GJ~357 is significantly brighter than the host stars of these other well constrained planetary systems, which offers significant advantages for future understanding of planetary dynamics and atmospheric characterisation. Detailed analyses of the planetary properties with additional observations, whilst searching for more companions on longer period orbits, or hidden worlds in the stable dynamical cavity between GJ~357b and d, can be made without large investments of telescope time.  This system provides us with a benchmark laboratory to understand the formation and evolutionary processes of small planets orbiting small stars.

\section*{Acknowledgments}
We thank the referee Stephen Kane for his detailed review of the manuscript that helped tighten-up various sections of our work.
JSJ and MT acknowledges funding by Fondecyt through grant 1161218 and partial support from CATA-Basal (PB06, Conicyt). ZMB acknowledges CONICYT-FONDECYT/Chile Postdoctorado 3180405. 
JIV acknowledges support of CONICYT-PFCHA/Doctorado Nacional-21191829, Chile.
MRD acknowledges support of CONICYT-PFCHA/Doctorado Nacional-21140646, Chile and Proyecto Basal AFB-170002. FJP and JCS acknowledges funding support from Spanish public funds for research under projects ESP2017-87676-C5-2-R.
\bibliographystyle{mnras}
\bibliography{refs}

\begin{thebibliography}{}
\makeatletter
\relax
\def\mn@urlcharsother{\let\do\@makeother \do\$\do\&\do\#\do\^\do\_\do\%\do\~}
\def\mn@doi{\begingroup\mn@urlcharsother \@ifnextchar [ {\mn@doi@}
  {\mn@doi@[]}}
\def\mn@doi@[#1]#2{\def\@tempa{#1}\ifx\@tempa\@empty \href
  {http://dx.doi.org/#2} {doi:#2}\else \href {http://dx.doi.org/#2} {#1}\fi
  \endgroup}
\def\mn@eprint#1#2{\mn@eprint@#1:#2::\@nil}
\def\mn@eprint@arXiv#1{\href {http://arxiv.org/abs/#1} {{\tt arXiv:#1}}}
\def\mn@eprint@dblp#1{\href {http://dblp.uni-trier.de/rec/bibtex/#1.xml}
  {dblp:#1}}
\def\mn@eprint@#1:#2:#3:#4\@nil{\def\@tempa {#1}\def\@tempb {#2}\def\@tempc
  {#3}\ifx \@tempc \@empty \let \@tempc \@tempb \let \@tempb \@tempa \fi \ifx
  \@tempb \@empty \def\@tempb {arXiv}\fi \@ifundefined
  {mn@eprint@\@tempb}{\@tempb:\@tempc}{\expandafter \expandafter \csname
  mn@eprint@\@tempb\endcsname \expandafter{\@tempc}}}

\bibitem[\protect\citeauthoryear{{Adibekyan}, {Sousa}, {Santos}, {Delgado
  Mena}, {Gonz{\'a}lez Hern{\'a}ndez}, {Israelian}, {Mayor}  \&
  {Khachatryan}}{{Adibekyan} et~al.}{2012}]{adibekyan12}
{Adibekyan} V.~Z.,  {Sousa} S.~G.,  {Santos} N.~C.,  {Delgado Mena} E.,
  {Gonz{\'a}lez Hern{\'a}ndez} J.~I.,  {Israelian} G.,  {Mayor} M.,
  {Khachatryan} G.,  2012, \mn@doi [A\&A] {10.1051/0004-6361/201219401}, \href
  {https://ui.adsabs.harvard.edu/abs/2012A%26A...545A..32A} {545, A32}

\bibitem[\protect\citeauthoryear{{Alberti}, {Carbone}, {Lepreti}  \&
  {Vecchio}}{{Alberti} et~al.}{2017}]{alberti17}
{Alberti} T.,  {Carbone} V.,  {Lepreti} F.,   {Vecchio} A.,  2017, \mn@doi
  [ApJ] {10.3847/1538-4357/aa78a2}, \href
  {https://ui.adsabs.harvard.edu/abs/2017ApJ...844...19A} {844, 19}

\bibitem[\protect\citeauthoryear{{Anglada-Escud{\'e}} \&
  {Butler}}{{Anglada-Escud{\'e}} \& {Butler}}{2012}]{AngladaButler2012}
{Anglada-Escud{\'e}} G.,  {Butler} R.~P.,  2012, \mn@doi [\apjs]
  {10.1088/0067-0049/200/2/15}, \href
  {http://adsabs.harvard.edu/abs/2012ApJS..200...15A} {200, 15}

\bibitem[\protect\citeauthoryear{{Anglada-Escud{\'e}}
  et~al.,}{{Anglada-Escud{\'e}} et~al.}{2013}]{anglada-escude13}
{Anglada-Escud{\'e}} G.,  et~al., 2013, \mn@doi [A\&A]
  {10.1051/0004-6361/201321331}, \href
  {http://adsabs.harvard.edu/abs/2013A%26A...556A.126A} {556, A126}

\bibitem[\protect\citeauthoryear{{Anglada-Escud{\'e}}
  et~al.,}{{Anglada-Escud{\'e}} et~al.}{2016}]{anglada-escude16}
{Anglada-Escud{\'e}} G.,  et~al., 2016, \mn@doi [Nature] {10.1038/nature19106},
  \href {https://ui.adsabs.harvard.edu/abs/2016Natur.536..437A} {536, 437}

\bibitem[\protect\citeauthoryear{{Astudillo-Defru}, {Delfosse}, {Bonfils},
  {Forveille}, {Lovis}  \& {Rameau}}{{Astudillo-Defru} et~al.}{2017}]{defru17}
{Astudillo-Defru} N.,  {Delfosse} X.,  {Bonfils} X.,  {Forveille} T.,  {Lovis}
  C.,   {Rameau} J.,  2017, \mn@doi [A\&A] {10.1051/0004-6361/201527078}, \href
  {https://ui.adsabs.harvard.edu/abs/2017A&A...600A..13A} {600, A13}

\bibitem[\protect\citeauthoryear{{Ballard}}{{Ballard}}{2019}]{ballard19}
{Ballard} S.,  2019, \mn@doi [\aj] {10.3847/1538-3881/aaf477}, \href
  {https://ui.adsabs.harvard.edu/abs/2019AJ....157..113B} {157, 113}

\bibitem[\protect\citeauthoryear{{Berdi{\~n}as}, {Rodr{\'{\i}}guez-L{\'o}pez},
  {Amado}, {Anglada-Escud{\'e}}, {Barnes}, {MacDonald}, {Zechmeister}  \&
  {Sarmiento}}{{Berdi{\~n}as} et~al.}{2017}]{berdinas17}
{Berdi{\~n}as} Z.~M.,  {Rodr{\'{\i}}guez-L{\'o}pez} C.,  {Amado} P.~J.,
  {Anglada-Escud{\'e}} G.,  {Barnes} J.~R.,  {MacDonald} J.,  {Zechmeister} M.,
    {Sarmiento} L.~F.,  2017, \mn@doi [MNRAS] {10.1093/mnras/stx1140}, \href
  {https://ui.adsabs.harvard.edu/abs/2017MNRAS.469.4268B} {469, 4268}

\bibitem[\protect\citeauthoryear{{Bonfils} et~al.,}{{Bonfils}
  et~al.}{2005}]{bonfils05a}
{Bonfils} X.,  et~al., 2005, \mn@doi [A\&A] {10.1051/0004-6361:200500193},
  \href {http://adsabs.harvard.edu/abs/2005A%26A...443L..15B} {443, L15}

\bibitem[\protect\citeauthoryear{{Bonfils} et~al.,}{{Bonfils}
  et~al.}{2013}]{bonfils2013}
{Bonfils} X.,  et~al., 2013, \mn@doi [\aap] {10.1051/0004-6361/201014704},
  \href {https://ui.adsabs.harvard.edu/abs/2013A&A...549A.109B} {549, A109}

\bibitem[\protect\citeauthoryear{{Borucki} et~al.,}{{Borucki}
  et~al.}{2010}]{borucki10}
{Borucki} W.~J.,  et~al., 2010, \mn@doi [Science] {10.1126/science.1185402},
  \href {https://ui.adsabs.harvard.edu/abs/2010Sci...327..977B} {327, 977}

\bibitem[\protect\citeauthoryear{{Buchner} et~al.,}{{Buchner}
  et~al.}{2014}]{PyMultiNest}
{Buchner} J.,  et~al., 2014, \mn@doi [\aap] {10.1051/0004-6361/201322971},
  \href {http://adsabs.harvard.edu/abs/2014A%26A...564A.125B} {564, A125}

\bibitem[\protect\citeauthoryear{{Burdanov} et~al.,}{{Burdanov}
  et~al.}{2019}]{burdanov19}
{Burdanov} A.~Y.,  et~al., 2019, arXiv e-prints, \href
  {https://ui.adsabs.harvard.edu/abs/2019arXiv190506035B} {}

\bibitem[\protect\citeauthoryear{{Butler} et~al.,}{{Butler}
  et~al.}{2017}]{butler2017}
{Butler} R.~P.,  et~al., 2017, \mn@doi [\aj] {10.3847/1538-3881/aa66ca}, \href
  {http://adsabs.harvard.edu/abs/2017AJ....153..208B} {153, 208}

\bibitem[\protect\citeauthoryear{{Cincotta} \& {Sim{\'o}}}{{Cincotta} \&
  {Sim{\'o}}}{1999}]{cincottasimo1999}
{Cincotta} P.,  {Sim{\'o}} C.,  1999, \mn@doi [Celestial Mechanics and
  Dynamical Astronomy] {10.1023/A:1008355215603}, \href
  {https://ui.adsabs.harvard.edu/\#abs/1999CeMDA..73..195C} {73, 195}

\bibitem[\protect\citeauthoryear{{Cincotta} \& {Sim{\'o}}}{{Cincotta} \&
  {Sim{\'o}}}{2000}]{cincottasimo2000}
{Cincotta} P.~M.,  {Sim{\'o}} C.,  2000, \mn@doi [Astronomy and Astrophysics
  Supplement Series] {10.1051/aas:2000108}, \href
  {https://ui.adsabs.harvard.edu/\#abs/2000A\&AS..147..205C} {147, 205}

\bibitem[\protect\citeauthoryear{{Cincotta}, {Giordano}  \&
  {Sim{\'o}}}{{Cincotta} et~al.}{2003}]{cincotta2003}
{Cincotta} P.~M.,  {Giordano} C.~M.,   {Sim{\'o}} C.,  2003, \mn@doi [Physica D
  Nonlinear Phenomena] {10.1016/S0167-2789(03)00103-9}, \href
  {https://ui.adsabs.harvard.edu/\#abs/2003PhyD..182..151C} {182, 151}

\bibitem[\protect\citeauthoryear{{Contro}, {Horner}, {Wittenmyer}, {Marshall}
  \& {Hinse}}{{Contro} et~al.}{2016}]{contro2016}
{Contro} B.,  {Horner} J.,  {Wittenmyer} R.~A.,  {Marshall} J.~P.,   {Hinse}
  T.~C.,  2016, \mn@doi [\mnras] {10.1093/mnras/stw1935}, \href
  {https://ui.adsabs.harvard.edu/abs/2016MNRAS.463..191C} {463, 191}

\bibitem[\protect\citeauthoryear{{D{\'{\i}}az} et~al.,}{{D{\'{\i}}az}
  et~al.}{2018}]{diaz2018}
{D{\'{\i}}az} M.~R.,  et~al., 2018, \mn@doi [\aj] {10.3847/1538-3881/aaa896},
  \href {http://adsabs.harvard.edu/abs/2018AJ....155..126D} {155, 126}

\bibitem[\protect\citeauthoryear{{D{\'\i}ez Alonso} et~al.,}{{D{\'\i}ez Alonso}
  et~al.}{2018a}]{diez18a}
{D{\'\i}ez Alonso} E.,  et~al., 2018a, \mn@doi [MNRAS] {10.1093/mnrasl/sly040},
  \href {https://ui.adsabs.harvard.edu/abs/2018MNRAS.476L..50D} {476, L50}

\bibitem[\protect\citeauthoryear{{D{\'\i}ez Alonso} et~al.,}{{D{\'\i}ez Alonso}
  et~al.}{2018b}]{diez18b}
{D{\'\i}ez Alonso} E.,  et~al., 2018b, \mn@doi [MNRAS] {10.1093/mnrasl/sly102},
  \href {https://ui.adsabs.harvard.edu/abs/2018MNRAS.480L...1D} {480, L1}

\bibitem[\protect\citeauthoryear{{Dressing} \& {Charbonneau}}{{Dressing} \&
  {Charbonneau}}{2015}]{dressing15}
{Dressing} C.~D.,  {Charbonneau} D.,  2015, \mn@doi [ApJ]
  {10.1088/0004-637X/807/1/45}, \href
  {https://ui.adsabs.harvard.edu/abs/2015ApJ...807...45D} {807, 45}

\bibitem[\protect\citeauthoryear{{Ducrot} et~al.,}{{Ducrot}
  et~al.}{2018}]{ducrot18}
{Ducrot} E.,  et~al., 2018, \mn@doi [AJ] {10.3847/1538-3881/aade94}, \href
  {https://ui.adsabs.harvard.edu/abs/2018AJ....156..218D} {156, 218}

\bibitem[\protect\citeauthoryear{{Espinoza}, {Kossakowski}  \&
  {Brahm}}{{Espinoza} et~al.}{2018}]{juliet}
{Espinoza} N.,  {Kossakowski} D.,   {Brahm} R.,  2018, arXiv e-prints, \href
  {https://ui.adsabs.harvard.edu/\#abs/2018arXiv181208549E} {p.
  arXiv:1812.08549}

\bibitem[\protect\citeauthoryear{{Fabrycky} et~al.,}{{Fabrycky}
  et~al.}{2014}]{fabrycky2014}
{Fabrycky} D.~C.,  et~al., 2014, \mn@doi [\apj] {10.1088/0004-637X/790/2/146},
  \href {https://ui.adsabs.harvard.edu/abs/2014ApJ...790..146F} {790, 146}

\bibitem[\protect\citeauthoryear{{Feng}, {Tuomi}, {Jones}, {Butler}  \&
  {Vogt}}{{Feng} et~al.}{2016}]{feng2016}
{Feng} F.,  {Tuomi} M.,  {Jones} H.~R.~A.,  {Butler} R.~P.,   {Vogt} S.,  2016,
  \mn@doi [\mnras] {10.1093/mnras/stw1478}, \href
  {http://adsabs.harvard.edu/abs/2016MNRAS.461.2440F} {461, 2440}

\bibitem[\protect\citeauthoryear{{Feng}, {Tuomi}, {Jones}, {Barnes},
  {Anglada-Escud{\'e}}, {Vogt}  \& {Butler}}{{Feng} et~al.}{2017}]{feng2017}
{Feng} F.,  {Tuomi} M.,  {Jones} H.~R.~A.,  {Barnes} J.,  {Anglada-Escud{\'e}}
  G.,  {Vogt} S.~S.,   {Butler} R.~P.,  2017, \mn@doi [\aj]
  {10.3847/1538-3881/aa83b4}, \href
  {http://adsabs.harvard.edu/abs/2017AJ....154..135F} {154, 135}

\bibitem[\protect\citeauthoryear{{Feroz}, {Hobson}  \& {Bridges}}{{Feroz}
  et~al.}{2009}]{MultiNest}
{Feroz} F.,  {Hobson} M.~P.,   {Bridges} M.,  2009, \mn@doi [\mnras]
  {10.1111/j.1365-2966.2009.14548.x}, \href
  {http://adsabs.harvard.edu/abs/2009MNRAS.398.1601F} {398, 1601}

\bibitem[\protect\citeauthoryear{{Figueira} et~al.,}{{Figueira}
  et~al.}{2012}]{figueira2012}
{Figueira} P.,  et~al., 2012, \mn@doi [\aap] {10.1051/0004-6361/201219017},
  \href {https://ui.adsabs.harvard.edu/abs/2012A&A...541A.139F} {541, A139}

\bibitem[\protect\citeauthoryear{{Fischer} \& {Valenti}}{{Fischer} \&
  {Valenti}}{2005}]{fischer05}
{Fischer} D.~A.,  {Valenti} J.,  2005, \mn@doi [ApJ] {10.1086/428383}, \href
  {http://adsabs.harvard.edu/cgi-bin/nph-bib_query?bibcode=2005ApJ...622.1102F&db_key=AST}
  {622, 1102}

\bibitem[\protect\citeauthoryear{{Foreman-Mackey}, {Hogg}, {Lang}  \&
  {Goodman}}{{Foreman-Mackey} et~al.}{2013}]{foreman13}
{Foreman-Mackey} D.,  {Hogg} D.~W.,  {Lang} D.,   {Goodman} J.,  2013, \mn@doi
  [PASP] {10.1086/670067}, \href
  {http://adsabs.harvard.edu/abs/2013PASP..125..306F} {125, 306}

\bibitem[\protect\citeauthoryear{{Foreman-Mackey}, {Agol}, {Angus}  \&
  {Ambikasaran}}{{Foreman-Mackey} et~al.}{2017}]{celerite}
{Foreman-Mackey} D.,  {Agol} E.,  {Angus} R.,   {Ambikasaran} S.,  2017,
  \mn@doi [AJ] {10.3847/1538-3881/aa9332}, 154, 220

\bibitem[\protect\citeauthoryear{{Forveille} et~al.,}{{Forveille}
  et~al.}{2009}]{forveille09}
{Forveille} T.,  et~al., 2009, \mn@doi [A\&A] {10.1051/0004-6361:200810557},
  \href {http://adsabs.harvard.edu/abs/2009A%26A...493..645F} {493, 645}

\bibitem[\protect\citeauthoryear{{Fulton}, {Petigura}, {Blunt}  \&
  {Sinukoff}}{{Fulton} et~al.}{2018}]{radvel}
{Fulton} B.~J.,  {Petigura} E.~A.,  {Blunt} S.,   {Sinukoff} E.,  2018, \mn@doi
  [\pasp] {10.1088/1538-3873/aaaaa8}, \href
  {http://adsabs.harvard.edu/abs/2018PASP..130d4504F} {130, 044504}

\bibitem[\protect\citeauthoryear{{Gaia Collaboration} et~al.,}{{Gaia
  Collaboration} et~al.}{2016}]{gaia16}
{Gaia Collaboration} et~al., 2016, \mn@doi [A\&A]
  {10.1051/0004-6361/201629272}, \href
  {http://adsabs.harvard.edu/abs/2016A%26A...595A...1G} {595, A1}

\bibitem[\protect\citeauthoryear{{Gaia Collaboration} et~al.,}{{Gaia
  Collaboration} et~al.}{2018}]{gaia18}
{Gaia Collaboration} et~al., 2018, \mn@doi [] {10.1051/0004-6361/201833051},
  \href {http://adsabs.harvard.edu/abs/2018A%26A...616A...1G} {616, A1}

\bibitem[\protect\citeauthoryear{{Gillon} et~al.,}{{Gillon}
  et~al.}{2017}]{gillon17}
{Gillon} M.,  et~al., 2017, \mn@doi [Nature] {10.1038/nature21360}, \href
  {https://ui.adsabs.harvard.edu/abs/2017Natur.542..456G} {542, 456}

\bibitem[\protect\citeauthoryear{{G{\"u}nther} et~al.,}{{G{\"u}nther}
  et~al.}{2019}]{gunter2019}
{G{\"u}nther} M.~N.,  et~al., 2019, arXiv e-prints, \href
  {https://ui.adsabs.harvard.edu/abs/2019arXiv190306107G} {p. arXiv:1903.06107}

\bibitem[\protect\citeauthoryear{{Haario}, {Saksman}  \& {Tamminen}}{{Haario}
  et~al.}{2001}]{haario01}
{Haario} H.,  {Saksman} E.,   {Tamminen} J.,  2001, Bernoulli, 7, 223

\bibitem[\protect\citeauthoryear{{Haario}, {Laine}, {Mira}  \&
  {Saksman}}{{Haario} et~al.}{2006}]{haario06}
{Haario} H.,  {Laine} M.,  {Mira} A.,   {Saksman} E.,  2006, Statistics and
  Computing, 16, 339

\bibitem[\protect\citeauthoryear{Hastings}{Hastings}{1970}]{hastings1970}
Hastings W.~K.,  1970, Biometrika, 57, 97

\bibitem[\protect\citeauthoryear{{Hinse}, {Christou}, {Alvarellos}  \&
  {Go{\'z}dziewski}}{{Hinse} et~al.}{2010}]{hinse2010}
{Hinse} T.~C.,  {Christou} A.~A.,  {Alvarellos} J.~L.~A.,   {Go{\'z}dziewski}
  K.,  2010, \mn@doi [\mnras] {10.1111/j.1365-2966.2010.16307.x}, \href
  {https://ui.adsabs.harvard.edu/abs/2010MNRAS.404..837H} {404, 837}

\bibitem[\protect\citeauthoryear{{Hinse}, {Haghighipour}, {Kostov}  \&
  {Go{\'z}dziewski}}{{Hinse} et~al.}{2015}]{hinse2015}
{Hinse} T.~C.,  {Haghighipour} N.,  {Kostov} V.~B.,   {Go{\'z}dziewski} K.,
  2015, \mn@doi [\apj] {10.1088/0004-637X/799/1/88}, \href
  {https://ui.adsabs.harvard.edu/abs/2015ApJ...799...88H} {799, 88}

\bibitem[\protect\citeauthoryear{{Houdebine}, {Mullan}, {Bercu}, {Paletou}  \&
  {Gebran}}{{Houdebine} et~al.}{2017}]{houdebine17}
{Houdebine} E.~R.,  {Mullan} D.~J.,  {Bercu} B.,  {Paletou} F.,   {Gebran} M.,
  2017, \mn@doi [ApJ] {10.3847/1538-4357/aa5cad}, \href
  {http://adsabs.harvard.edu/abs/2017ApJ...837...96H} {837, 96}

\bibitem[\protect\citeauthoryear{{Jenkins}, {Jones}, {Pavlenko}, {Pinfield},
  {Barnes}  \& {Lyubchik}}{{Jenkins} et~al.}{2008}]{jenkins08}
{Jenkins} J.~S.,  {Jones} H.~R.~A.,  {Pavlenko} Y.,  {Pinfield} D.~J.,
  {Barnes} J.~R.,   {Lyubchik} Y.,  2008, \mn@doi [A\&A]
  {10.1051/0004-6361:20078611}, \href
  {http://adsabs.harvard.edu/abs/2008A%26A...485..571J} {485, 571}

\bibitem[\protect\citeauthoryear{{Jenkins} et~al.,}{{Jenkins}
  et~al.}{2009}]{jenkins09}
{Jenkins} J.~S.,  et~al., 2009, \mn@doi [MNRAS]
  {10.1111/j.1365-2966.2009.15097.x}, \href
  {http://adsabs.harvard.edu/abs/2009MNRAS.398..911J} {398, 911}

\bibitem[\protect\citeauthoryear{{Jenkins} et~al.,}{{Jenkins}
  et~al.}{2011}]{jenkins11a}
{Jenkins} J.~S.,  et~al., 2011, \mn@doi [A\&A] {10.1051/0004-6361/201016333},
  \href {http://adsabs.harvard.edu/abs/2011A\%26A...531A...8J} {531, A8}

\bibitem[\protect\citeauthoryear{{Jenkins} et~al.,}{{Jenkins}
  et~al.}{2016}]{jenkinsjm16}
{Jenkins} J.~M.,  et~al., 2016, in Software and Cyberinfrastructure for
  Astronomy IV. p. 99133E, \mn@doi{10.1117/12.2233418}

\bibitem[\protect\citeauthoryear{{Jenkins} et~al.,}{{Jenkins}
  et~al.}{2017}]{jenkins17}
{Jenkins} J.~S.,  et~al., 2017, \mn@doi [MNRAS] {10.1093/mnras/stw2811}, \href
  {https://ui.adsabs.harvard.edu/abs/2017MNRAS.466..443J} {466, 443}

\bibitem[\protect\citeauthoryear{{Johnson}, {Aller}, {Howard}  \&
  {Crepp}}{{Johnson} et~al.}{2010}]{johnson10}
{Johnson} J.~A.,  {Aller} K.~M.,  {Howard} A.~W.,   {Crepp} J.~R.,  2010,
  \mn@doi [PASP] {10.1086/655775}, \href
  {https://ui.adsabs.harvard.edu/abs/2010PASP..122..905J} {122, 905}

\bibitem[\protect\citeauthoryear{{Jontof-Hutter} et~al.,}{{Jontof-Hutter}
  et~al.}{2016}]{jontof16}
{Jontof-Hutter} D.,  et~al., 2016, \mn@doi [ApJ] {10.3847/0004-637X/820/1/39},
  \href {https://ui.adsabs.harvard.edu/abs/2016ApJ...820...39J} {820, 39}

\bibitem[\protect\citeauthoryear{{Kass} \& {Raftery}}{{Kass} \&
  {Raftery}}{1995}]{kass95}
{Kass} R.~E.,  {Raftery} A.~E.,  1995, J. Am. Stat. Ass., 430, 773

\bibitem[\protect\citeauthoryear{{Kopparapu}}{{Kopparapu}}{2013}]{kopparapu13}
{Kopparapu} R.~K.,  2013, \mn@doi [ApJL] {10.1088/2041-8205/767/1/L8}, \href
  {https://ui.adsabs.harvard.edu/abs/2013ApJ...767L...8K} {767, L8}

\bibitem[\protect\citeauthoryear{{Kreidberg}}{{Kreidberg}}{2015}]{batman}
{Kreidberg} L.,  2015, \mn@doi [Publications of the Astronomical Society of the
  Pacific] {10.1086/683602}, \href
  {https://ui.adsabs.harvard.edu/\#abs/2015PASP..127.1161K} {127, 1161}

\bibitem[\protect\citeauthoryear{{Laughlin}, {Bodenheimer}  \&
  {Adams}}{{Laughlin} et~al.}{2004}]{laughlin04a}
{Laughlin} G.,  {Bodenheimer} P.,   {Adams} F.~C.,  2004, ApJL, \href
  {http://adsabs.harvard.edu/cgi-bin/nph-bib_query?bibcode=2004ApJ...612L..73L&amp;db_key=AST}
  {612, L73}

\bibitem[\protect\citeauthoryear{{Lee}, {Peale}, {Pfahl}  \& {Ward}}{{Lee}
  et~al.}{2007}]{lee2007}
{Lee} M.~H.,  {Peale} S.~J.,  {Pfahl} E.,   {Ward} W.~R.,  2007, \mn@doi
  [\icarus] {10.1016/j.icarus.2007.03.005}, \href
  {https://ui.adsabs.harvard.edu/abs/2007Icar..190..103L} {190, 103}

\bibitem[\protect\citeauthoryear{{Libert} \& {Tsiganis}}{{Libert} \&
  {Tsiganis}}{2009a}]{libert2009}
{Libert} A.~S.,  {Tsiganis} K.,  2009a, \mn@doi [\mnras]
  {10.1111/j.1365-2966.2009.15532.x}, \href
  {https://ui.adsabs.harvard.edu/abs/2009MNRAS.400.1373L} {400, 1373}

\bibitem[\protect\citeauthoryear{{Libert} \& {Tsiganis}}{{Libert} \&
  {Tsiganis}}{2009b}]{libertA2009}
{Libert} A.~S.,  {Tsiganis} K.,  2009b, \mn@doi [\aap]
  {10.1051/0004-6361:200810843}, \href
  {https://ui.adsabs.harvard.edu/abs/2009A\&A...493..677L} {493, 677}

\bibitem[\protect\citeauthoryear{{Liddle}}{{Liddle}}{2007}]{liddle2007}
{Liddle} A.~R.,  2007, \mn@doi [\mnras] {10.1111/j.1745-3933.2007.00306.x},
  \href {http://adsabs.harvard.edu/abs/2007MNRAS.377L..74L} {377, L74}

\bibitem[\protect\citeauthoryear{{Luque} et~al.,}{{Luque}
  et~al.}{2019}]{luque19}
{Luque} R.,  et~al., 2019, arXiv e-prints, \href
  {https://ui.adsabs.harvard.edu/abs/2019arXiv190412818L} {}

\bibitem[\protect\citeauthoryear{{MacDonald} et~al.,}{{MacDonald}
  et~al.}{2016}]{macdonald16}
{MacDonald} M.~G.,  et~al., 2016, \mn@doi [AJ] {10.3847/0004-6256/152/4/105},
  \href {https://ui.adsabs.harvard.edu/abs/2016AJ....152..105M} {152, 105}

\bibitem[\protect\citeauthoryear{{Mann} et~al.,}{{Mann} et~al.}{2017}]{mann17}
{Mann} A.~W.,  et~al., 2017, \mn@doi [AJ] {10.3847/1538-3881/aa7140}, \href
  {https://ui.adsabs.harvard.edu/abs/2017AJ....153..267M} {153, 267}

\bibitem[\protect\citeauthoryear{{Mayor} et~al.,}{{Mayor}
  et~al.}{2009}]{mayor09b}
{Mayor} M.,  et~al., 2009, \mn@doi [A\&A] {10.1051/0004-6361/200912172}, \href
  {http://adsabs.harvard.edu/abs/2009A%26A...507..487M} {507, 487}

\bibitem[\protect\citeauthoryear{{Metropolis}, {Rosenbluth}, {Rosenbluth},
  {Teller}  \& {Teller}}{{Metropolis} et~al.}{1953}]{metropolis1953}
{Metropolis} N.,  {Rosenbluth} A.~W.,  {Rosenbluth} M.~N.,  {Teller} A.~H.,
  {Teller} E.,  1953, \mn@doi [\jcp] {10.1063/1.1699114}, \href
  {https://ui.adsabs.harvard.edu/abs/1953JChPh..21.1087M} {21, 1087}

\bibitem[\protect\citeauthoryear{{Miles-P{\'a}ez}, {Zapatero Osorio},
  {Pall{\'e}}  \& {Metchev}}{{Miles-P{\'a}ez} et~al.}{2019}]{miles-paez19}
{Miles-P{\'a}ez} P.~A.,  {Zapatero Osorio} M.~R.,  {Pall{\'e}} E.,   {Metchev}
  S.~A.,  2019, \mn@doi [MNRAS] {10.1093/mnrasl/slz001}, \href
  {https://ui.adsabs.harvard.edu/abs/2019MNRAS.484L..38M} {484, L38}

\bibitem[\protect\citeauthoryear{{Millholland} et~al.,}{{Millholland}
  et~al.}{2018}]{millho2018}
{Millholland} S.,  et~al., 2018, \mn@doi [\aj] {10.3847/1538-3881/aaa894},
  \href {https://ui.adsabs.harvard.edu/abs/2018AJ....155..106M} {155, 106}

\bibitem[\protect\citeauthoryear{{Montes} et~al.,}{{Montes}
  et~al.}{2018}]{montes18}
{Montes} D.,  et~al., 2018, \mn@doi [MNRAS] {10.1093/mnras/sty1295}, \href
  {https://ui.adsabs.harvard.edu/abs/2018MNRAS.479.1332M} {479, 1332}

\bibitem[\protect\citeauthoryear{{Moran}, {H{\"o}rst}, {Batalha}, {Lewis}  \&
  {Wakeford}}{{Moran} et~al.}{2018}]{moran18}
{Moran} S.~E.,  {H{\"o}rst} S.~M.,  {Batalha} N.~E.,  {Lewis} N.~K.,
  {Wakeford} H.~R.,  2018, \mn@doi [AJ] {10.3847/1538-3881/aae83a}, \href
  {https://ui.adsabs.harvard.edu/abs/2018AJ....156..252M} {156, 252}

\bibitem[\protect\citeauthoryear{{Muirhead} et~al.,}{{Muirhead}
  et~al.}{2015}]{muirhead15}
{Muirhead} P.~S.,  et~al., 2015, \mn@doi [ApJ] {10.1088/0004-637X/801/1/18},
  \href {https://ui.adsabs.harvard.edu/abs/2015ApJ...801...18M} {801, 18}

\bibitem[\protect\citeauthoryear{{Muirhead}, {Dressing}, {Mann}, {Rojas-Ayala},
  {L{\'e}pine}, {Paegert}, {De Lee}  \& {Oelkers}}{{Muirhead}
  et~al.}{2018}]{muirhead18}
{Muirhead} P.~S.,  {Dressing} C.~D.,  {Mann} A.~W.,  {Rojas-Ayala} B.,
  {L{\'e}pine} S.,  {Paegert} M.,  {De Lee} N.,   {Oelkers} R.,  2018, \mn@doi
  [AJ] {10.3847/1538-3881/aab710}, \href
  {http://adsabs.harvard.edu/abs/2018AJ....155..180M} {155, 180}

\bibitem[\protect\citeauthoryear{{Neves}, {Bonfils}, {Santos}, {Delfosse},
  {Forveille}, {Allard}  \& {Udry}}{{Neves} et~al.}{2014}]{neves14}
{Neves} V.,  {Bonfils} X.,  {Santos} N.~C.,  {Delfosse} X.,  {Forveille} T.,
  {Allard} F.,   {Udry} S.,  2014, \mn@doi [A\&A]
  {10.1051/0004-6361/201424139}, \href
  {http://adsabs.harvard.edu/abs/2014A%26A...568A.121N} {568, A121}

\bibitem[\protect\citeauthoryear{{Pe\~na Rojas} \& {Jenkins}}{{Pe\~na Rojas} \&
  {Jenkins}}{2019}]{pena19}
{Pe\~na Rojas} P.~A.,  {Jenkins} J.~S.,  2019, A\&A, p. in prep

\bibitem[\protect\citeauthoryear{{Perryman}, {Lindegren}, {Kovalevsky}  \& {et
  al.}}{{Perryman} et~al.}{1997}]{perryman97}
{Perryman} M.~A.~C.,  {Lindegren} L.,  {Kovalevsky} J.,   {et al.} 1997, A\&A,
  \href
  {http://adsabs.harvard.edu/cgi-bin/nph-bib_query?bibcode=1997A%26A...323L..49P&db_key=AST}
  {323, L49}

\bibitem[\protect\citeauthoryear{{Pojmanski}}{{Pojmanski}}{1997}]{pojmanski97}
{Pojmanski} G.,  1997, Acta Astronomica, \href
  {http://adsabs.harvard.edu/abs/1997AcA....47..467P} {47, 467}

\bibitem[\protect\citeauthoryear{{Rein} \& {Liu}}{{Rein} \&
  {Liu}}{2012}]{rein2012}
{Rein} H.,  {Liu} S.-F.,  2012, \mn@doi [\aap] {10.1051/0004-6361/201118085},
  \href {http://adsabs.harvard.edu/abs/2012A\%26A...537A.128R} {537, A128}

\bibitem[\protect\citeauthoryear{{Rein} \& {Tamayo}}{{Rein} \&
  {Tamayo}}{2015}]{rein2015}
{Rein} H.,  {Tamayo} D.,  2015, \mn@doi [\mnras] {10.1093/mnras/stv1257}, \href
  {http://adsabs.harvard.edu/abs/2015MNRAS.452..376R} {452, 376}

\bibitem[\protect\citeauthoryear{{Ribas} et~al.,}{{Ribas}
  et~al.}{2018}]{ribas18}
{Ribas} I.,  et~al., 2018, \mn@doi [Nature] {10.1038/s41586-018-0677-y}, \href
  {https://ui.adsabs.harvard.edu/abs/2018Natur.563..365R} {563, 365}

\bibitem[\protect\citeauthoryear{{Ricker} et~al.,}{{Ricker}
  et~al.}{2015}]{TESS}
{Ricker} G.~R.,  et~al., 2015, \mn@doi [Journal of Astronomical Telescopes,
  Instruments, and Systems] {10.1117/1.JATIS.1.1.014003}, \href
  {http://adsabs.harvard.edu/abs/2015JATIS...1a4003R} {1, 014003}

\bibitem[\protect\citeauthoryear{{Rojas-Ayala}, {Covey}, {Muirhead}  \&
  {Lloyd}}{{Rojas-Ayala} et~al.}{2010}]{rojas-ayala10}
{Rojas-Ayala} B.,  {Covey} K.~R.,  {Muirhead} P.~S.,   {Lloyd} J.~P.,  2010,
  \mn@doi [ApJL] {10.1088/2041-8205/720/1/L113}, \href
  {https://ui.adsabs.harvard.edu/abs/2010ApJ...720L.113R} {720, L113}

\bibitem[\protect\citeauthoryear{{Savitzky} \& {Golay}}{{Savitzky} \&
  {Golay}}{1964}]{savitzkygolay}
{Savitzky} A.,  {Golay} M.~J.~E.,  1964, Analytical Chemistry, \href
  {http://adsabs.harvard.edu/abs/1964AnaCh..36.1627S} {36, 1627}

\bibitem[\protect\citeauthoryear{{Schlaufman} \& {Laughlin}}{{Schlaufman} \&
  {Laughlin}}{2010}]{schlaufman10}
{Schlaufman} K.~C.,  {Laughlin} G.,  2010, \mn@doi [A\&A]
  {10.1051/0004-6361/201015016}, \href
  {http://cdsads.u-strasbg.fr/abs/2010A%26A...519A.105S} {519, A105}

\bibitem[\protect\citeauthoryear{{Sch{\"o}fer} et~al.,}{{Sch{\"o}fer}
  et~al.}{2019}]{schofer19}
{Sch{\"o}fer} P.,  et~al., 2019, \mn@doi [A\&A] {10.1051/0004-6361/201834114},
  \href {https://ui.adsabs.harvard.edu/abs/2019A&A...623A..44S} {623, A44}

\bibitem[\protect\citeauthoryear{{Sousa}, {Santos}, {Israelian}, {Mayor}  \&
  {Udry}}{{Sousa} et~al.}{2011}]{sousa11}
{Sousa} S.~G.,  {Santos} N.~C.,  {Israelian} G.,  {Mayor} M.,   {Udry} S.,
  2011, \mn@doi [A\&A] {10.1051/0004-6361/201117699}, \href
  {http://adsabs.harvard.edu/abs/2011A%26A...533A.141S} {533, A141}

\bibitem[\protect\citeauthoryear{{Stassun} et~al.,}{{Stassun}
  et~al.}{2018}]{stassun18}
{Stassun} K.~G.,  et~al., 2018, \mn@doi [AJ] {10.3847/1538-3881/aad050}, \href
  {http://adsabs.harvard.edu/abs/2018AJ....156..102S} {156, 102}

\bibitem[\protect\citeauthoryear{{Steffen} et~al.,}{{Steffen}
  et~al.}{2013}]{steffen13}
{Steffen} J.~H.,  et~al., 2013, \mn@doi [MNRAS] {10.1093/mnras/sts090}, \href
  {https://ui.adsabs.harvard.edu/abs/2013MNRAS.428.1077S} {428, 1077}

\bibitem[\protect\citeauthoryear{{Tremaine} \& {Dong}}{{Tremaine} \&
  {Dong}}{2012}]{tremaine2012}
{Tremaine} S.,  {Dong} S.,  2012, \mn@doi [\aj] {10.1088/0004-6256/143/4/94},
  \href {https://ui.adsabs.harvard.edu/abs/2012AJ....143...94T} {143, 94}

\bibitem[\protect\citeauthoryear{{Tuomi}}{{Tuomi}}{2011}]{tuomi11b}
{Tuomi} M.,  2011, \mn@doi [A\&A] {10.1051/0004-6361/201015995}, \href
  {http://adsabs.harvard.edu/abs/2011A%26A...528L...5T} {528, L5}

\bibitem[\protect\citeauthoryear{{Tuomi}, {Jones}, {Barnes},
  {Anglada-Escud{\'e}}  \& {Jenkins}}{{Tuomi} et~al.}{2014}]{tuomi14b}
{Tuomi} M.,  {Jones} H.~R.~A.,  {Barnes} J.~R.,  {Anglada-Escud{\'e}} G.,
  {Jenkins} J.~S.,  2014, \mn@doi [MNRAS] {10.1093/mnras/stu358}, \href
  {http://adsabs.harvard.edu/abs/2014MNRAS.441.1545T} {441, 1545}

\bibitem[\protect\citeauthoryear{{Tuomi}, {Jones}, {Barnes},
  {Anglada-Escud{\'e}}, {Butler}, {Kiraga}  \& {Vogt}}{{Tuomi}
  et~al.}{2018}]{tuomi2018}
{Tuomi} M.,  {Jones} H.~R.~A.,  {Barnes} J.~R.,  {Anglada-Escud{\'e}} G.,
  {Butler} R.~P.,  {Kiraga} M.,   {Vogt} S.~S.,  2018, \mn@doi [\aj]
  {10.3847/1538-3881/aab09c}, \href
  {http://adsabs.harvard.edu/abs/2018AJ....155..192T} {155, 192}

\bibitem[\protect\citeauthoryear{{Tuomi}, {Jones}  \&
  {Anglada-Escud{\'e}}}{{Tuomi} et~al.}{2019}]{tuomi2019}
{Tuomi} M.,  {Jones} H.~R.~A.,   {Anglada-Escud{\'e}} G. e.~a.,  2019, \apj

\bibitem[\protect\citeauthoryear{{Winn}}{{Winn}}{2010}]{josua2010}
{Winn} J.~N.,  2010, {Exoplanet Transits and Occultations}.
pp 55--77

\bibitem[\protect\citeauthoryear{{Zechmeister} \& {K{\"u}rster}}{{Zechmeister}
  \& {K{\"u}rster}}{2009}]{zechmeister09b}
{Zechmeister} M.,  {K{\"u}rster} M.,  2009, \mn@doi [A\&A]
  {10.1051/0004-6361:200811296}, \href
  {https://ui.adsabs.harvard.edu/abs/2009A&A...496..577Z} {496, 577}

\bibitem[\protect\citeauthoryear{{Zechmeister}, {K{\"u}rster}  \&
  {Endl}}{{Zechmeister} et~al.}{2009}]{zechmeister2009}
{Zechmeister} M.,  {K{\"u}rster} M.,   {Endl} M.,  2009, \mn@doi [\aap]
  {10.1051/0004-6361/200912479}, \href
  {https://ui.adsabs.harvard.edu/abs/2009A&A...505..859Z} {505, 859}

\makeatother
\end{thebibliography}

\end{document}